\documentclass[aps,prd,preprintnumbers,showpacs,showkeys,nofootinbib,
superscriptaddress,fleqn,floatfix, tightenlines]{revtex4-1}
\usepackage{amsmath,amsfonts,amssymb,amscd,amsxtra,amsthm}
\usepackage{graphicx}
\usepackage{epstopdf}
\usepackage{bm}
\usepackage{subfigure}

\usepackage{slashed}
\usepackage[utf8]{inputenc}
\usepackage[normalem]{ulem} 
\usepackage[dvipsnames]{xcolor} 
\renewcommand\sout{\bgroup \color{red} \ULdepth=-.5ex \ULset}

\date{\today}

\makeatother

\begin{document}
\preprint{INHA-NTG-04/2016}
\title{Photoproduction of the scalar mesons $f_0(500)$ and $f_0(980)$
  off the nucleon}
\author{Je-Hee Lee}
\email{jehee.lee@inha.edu}
\affiliation{Department of Physics, Inha University,
Incheon 22212, Republic of Korea}
\affiliation{RIKEN Nishina Center, RIKEN, 2-1 Hirosawa, 351-0115
  Saitama, Japan}

\author{Hyun-Chul Kim}
\email{hchkim@inha.ac.kr}
\affiliation{Department of Physics, Inha University,
Incheon 22212, Republic of Korea}
\affiliation{School of Physics, Korea Institute for Advanced Study
  (KIAS), Seoul 02455, Republic of Korea}

\author{Sang-Ho Kim}
\email{sangho.kim@apctp.org}
\affiliation{Asia Pacific Center for Theoretical Physics (APCTP), Pohang
37673, Republic of Korea}

\author{Hui-Young Ryu}
\email{hyryu2015@knu.ac.kr}
\affiliation{Department of Physics, Pusan National University, Busan
46241, Republic of Korea}

\author{Byung-Geel Yu}
\email{bgyu@kau.ac.kr}
\affiliation{Research Institute for Basic Sciences, Korea Aerospace
  University, Goyang 10540, Republic of Korea}

\begin{abstract}
We investigate photoproduction of the scalar mesons off the nucleon,
using the effective Lagrangians and the Regge approach. We first study
$f_0(980)$ photoproduction, replacing the Feynman propagator with the
Regge ones for $\rho$-meson exchange in the $t$-channel. The
model parameters are fixed by reproducing the experimental data on the
differential cross section. We then apply the same method to
$f_0(500)$ or $\sigma$ photoproduction with the same parameters for
the Regge propagator. Since the threshold energy of the $f_0(500)$
production is rather low, $N^*$ resonances, which can decay into the
nucleon and two pions in the isoscalar and scalar state, can come into
play in the $s$-channel. To examine the effects of the $N^*$
resonances, we set up two models to look into the respective
contribution of the $N^*$ resonances. We find that in particular
$N(1440)$ and $N(1680)$ play important roles in $f_0(500)$
photoproduction. We discuss the physical implications of the present
results.
\end{abstract} 


\maketitle

\section{Introduction}
Understanding the structure of low-lying scalar mesons has been one of
the most challenging issues in hadronic physics. Their internal
structure is still under debate. That the $f_0(500)$ scalar meson, which
is also known as $\sigma$, is not an ordinary meson consisting of
a quark and an anti-quark is more or less in consensus. 
Recent studies suggest that these scalar mesons may belong to the
flavor SU(3) non-$q\bar{q}$ nonet~(see
reviews~\cite{Amsler:2004ps,Bugg:2004xu}, a
``\emph{note on scalar mesons below 2 GeV}'' in
Ref.~\cite{PDG2016}, and references therein. 
 A recent review provides also various information on the 
 structure of the scalar meson~\cite{Pelaez:2015qba}, including a
 historical background of the $\sigma$ meson). 
The $f_0(500)$ is also interepreted as one of the glueballs or
gluonia, mixed with the $\bar{q}q$ state~\cite{Minkowski:1998mf,
  Ochs:2013gi,Mennessier:2010xg}, though this idea is criticized
because the same analysis is rather difficult to be applied to
explaining the strange scalar meson $K_0^*(800)$ or $\kappa$, which is
also considered as a member of the nonet. The $f_0(500)$ is often
regarded as a tetraquark state in a broad sense~\cite{Jaffe:1976ig}. 
The $f_0(500)$ as a tetraquark state has a multiple meaning: It can be
described as a diquark-antidiquark correlated
state~\cite{Maiani:2004uc, Hooft:2008we}, $\bar{q}q\bar{q}q$
state~\cite{Eichmann:2015cra}, or correlated $2\pi$  
state~\cite{Lohse:1990ew, Oller:1998hw} arising from $\pi\pi$
scattering. This non $\bar{q} q$ feature was employed in various
theoretical approaches such as QCD sum rules~\cite{Chen:2007xr},
effective Lagrangians~\cite{Black:1998wt}, and lattice
QCD~\cite{Kunihiro:2003yj, Wakayama:2014gpa, Soto:2011ap}. 

The scalar mesons were also extensively studied
phenomenologically. There are two scalar-isoscalar mesons 
($I^G(J^{PC})=0^+(0^{++})$) below 1 GeV, that is, the lowest-lying
$f_0(500)$ (or $\sigma$) and the first excited $f_0(980)$. Both the
$f_0(500)$ and the $f_0(980)$ exist in $\pi\pi$ scattering and their
pole positions were investigated based on many  different processes,
for example, such as $\pi N\to \pi\pi N$ 
reactions~\cite{Hyams:1973zf, Protopopescu:1973sh, Grayer:1974cr},
$K_{l4}$ decay~\cite{Rosselet:1976pu, Batley:2010zza}, $D\to
3\pi$~\cite{Aitala:2000xu,Bonvicini:2007tc}, $J/\psi\to \omega
\pi\pi$~\cite{Ablikim:2004qna}, $\psi(2\mathrm{S})\to \pi^+\pi^-
J/\psi$~\cite{Ablikim:2006bz},
$\gamma\gamma\to\pi\pi$~\cite{Hoferichter:2011wk}, $pp$
scattering~\cite{Barberis:1999cq}, and so on (for details, we refer to
Refs.~\cite{PDG2016, Pelaez:2015qba}).  While the mass and the
width of the $f_0(980)$ are more or less known to be $m_{f_0}=990\pm
20$ MeV and $\Gamma=40-100$ MeV, those of $f_0(500)$ are still far
from consensus (see Ref.~~\cite{PDG2016}).  The upper
bound of the $f_0(500)$ mass is given in the large $N_c$ limit in
terms of the Gasser-Leutwyler low-energy
constant~\cite{Polyakov:2001ua}, which suggests that the $f_0$ mass is
quite possibly smaller than 700 MeV.

While there was a great deal of theoretical works on the structure of
the $f_0(500)$, its reacion mechanism was less investigated. 
Recently, the CLAS Collaboration has reported the first analysis of
the $S$-wave photoproduction of $\pi^+\pi^-$ pairs in the region of
the $f_0(980)$ at photon energies between 3.0 and 3.8 GeV and momentum
transfer squared $-t$ between $0.4\,\mathrm{GeV}^2$ and
$1\,\mathrm{GeV}^2$~\cite{Battaglieri:2008ps,
  Battaglieri:2009aa}. While the differential cross section for the
$\gamma p\to \pi^+\pi^-p$ process in the $S$-wave shows an evident
signal for the $f_0(980)$ production, the $f_0(500)$ was not seen
clearly. However, there is still a hint for the existence of the
$f_0(500)$ in $\pi^+\pi^- p$ photoproduction measured at different
kinematic conditions~\cite{CLAS}. Thus, it is of great interest to
study the $\gamma p\to \pi^+\pi^- p$ reaction in the scalar and
isoscalar channel. Since these two pions are strongly correlated, one
has to consider the rescattering effects of these two pions to
describe $\gamma p\to \pi^+\pi^- p$ in the scalar and isoscalar
channel, which are essential in order to explain the production
mechanism of the scalar and isoscalar mesons $f_0$ quantitatively in
$\pi^+\pi^- p$ photoproduction. Moreover, it is crucial to take into
account the $K\bar{K}$ channel in addition~\cite{Ji:1997fb}, since its
threshold is open in the vicinity of the $f_0(980)$ mass.
In order to take into account the effects of the $K\bar{K}$ channel,
one has to introduce the coupled-channel formalism, which requires the
fully coupled $\pi\pi$ and $K\bar{K}$ amplitudes.

However, before we carry out the investigation on the $\gamma p\to
\pi^+\pi^- p$ reaction, we need to examine the related two-body
process $\gamma p\to f_0 p$ as a first step toward more complicated
correlated $\pi\pi$ photoproduction. Moreover, since the CLAS
Collaboration already presented the differential cross section for
$f_0(980)$ photoproduction in the photon energy range
$E_\gamma=(3.0-3.8)$ GeV, it is important to study $f_0$ photoproduction
theoretically as well before we examine the $\gamma p\to \pi^+\pi^- p$
process with pion pairs in the $S$-wave. In addition to $f_0(980)$
photoproduction, we study in the present work the $f_0(500)$
production by photon beams, based on effective Lagrangians and a Regge
approach. The Regge exchange in $f_0(980)$ photoproduction was already
applied in Ref.~\cite{Donnachie:2015jaa} with the same Regge
trajectory but a different set of parameters. We will first compute
the differential cross section for $f_0(980)$ photoproduction and
compare the results with the CLAS experimental data such that we can
fix parameters for the $t$-channel Reggeon exchange.
Since there is no excited nucleon that decays into $\pi^+\pi^-$ pairs
in the $S$-wave beyond the $f_0(980)N$ threshold, we consider only the
$N$ exchange in the $s$ channel. Then we will proceed to study
$f_0(500)$ photoproduction with the same parameters for the Reggeon,
which is fixed in $f_0(980)$ production. As far as $f_0(500)$
photoproduction is concerned, we need to consider several excited
nucleons above the threshold energy, which can decay into  $\pi^+\pi^-
N$, where the pion pairs are in the isoscalar and scalar wave.

Upon computing the transition amplitude for the $f_0(500)$
photoproduction, there are ambiguities to which we have to pay
attention carefully. Firstly, the width of the $f_0(500)$ is very
large, so that the value of the $f_0(500)$ mass is quite
uncertain. Thus, we have to look into the dependence of the results on
the $f_0(500)$ mass.
Secondly, the photocoupling constant $g_{\gamma\sigma\rho}$ is
experimentally not much known. Though there are several
theoretical suggestions on its value, the agreement has not been
reached yet. Experimentally, two relevant decay channels are known:
$\rho^0\to \pi^+\pi^-\gamma$~\cite{Dolinsky:1991vq,Vasserman:1988yr}
and $\rho^0\to \pi^0\pi^0\gamma$~\cite{Achasov:2002jv,Akhmetshin:2006bx}.
To see the contribution of the $f_0(500)$ in these decay processes,
one has to reply on models. So, it is required to examine
uncertainties arising from the coupling constant
$g_{\gamma\sigma\rho}$. In principle, $\omega$-meson exchange could be
considered. However, the branching ratio of $\omega \to \pi^+\pi^-
\gamma$ is not much known: its experimental upper bound is given as
$<3.6\times 10^{-3}$ with CL=95\,\%. On the other hand, $\rho\to
\pi^+\pi^-\gamma$ is experimentally known to be $(9.9\pm 1.6)\times
10^{-3}$~\cite{PDG2016}. Thus, the value of the $\gamma f_0
\rho$ coupling constant is expected to be much larger than that of
$\omega \to f_0 \gamma$, based on the experimental data given
above. We have confirmed numerically that the effect of
$\omega$-exchange is indeed much smaller than that of $\rho$-exchange.
So, we will ignore in this work the contribution from $\omega$-meson
exchange.
Thirdly, the final state in the $N^*\to
(\pi\pi)_{S-\mathrm{wave}}^{I=0} N$ decay should contain both the
background $\pi\pi$ and the $f_0(500)$ resonance. It indicates that it
is rather difficult to determine the coupling constants for the
$f_0(500)NN^*$ unambiguously. Considering these points that will
bring about the uncertainties of the present work, we have to
introduce certain assumptions before we proceed to investigate
$f_0(500)$ photoproduction. Though we will take 500 MeV as a main
value for the $f_0(500)$ mass in this work, we will carefully examine
the dependence of the results for the total cross section on the mass
of the lowest-lying scalar meson.
Lastly, we will regard $\pi^+\pi^-$ pairs in the $S$-wave as the
$f_0(500)$ meson, which are produced in the course of the $N^*\to
(\pi^+\pi^-)_{S-\mathrm{wave}}^{I=0} N$ decays, so that we are able to
determine the strong coupling constants for the $N^*\to f_0(500) N$
transitions. Since the $f_0(500)$ resonance is
the most dominant one in $\pi\pi$ scattering in the scalar-isoscalar
channel, this approximation is rather plausible.

In addition to the Roper resonance, we want to consider other $N^*$
resonances that can decay into
$(\pi\pi)_{S-\mathrm{wave}}^{I=0}N$. Referring to
Ref.~\cite{PDG2016}, we find that 10 excited nucleon
resonances have the decay channel of $N^* \to
(\pi\pi)_{S-\mathrm{wave}}^{I=0}N$. However, there are not enough data
for $N^*\to (\pi\pi)_{S-\mathrm{wave}}^{I=0}$ except for $N(1440)$,
$N(1680)$, and $N(1880)$. Since $N(1880)$ has an overall status 2
star, we will not consider it (see the review ``\textit{$N$ and $\Delta$
resonances}'' in Ref.~\cite{PDG2016}). Thus, we expect that the
main contribution will come from $N(1440)$ and $N(1680)$. As will see
later, $N(1680)$ provides predominantly a large contribution to the
total cross section. Thus, we will set up two different models to
delve into each contribution from the $N^*$ resonances. In Model I, we
will include those with spin $1/2$ and an overall status 3 or 4
stars. Thus, we consider $N(1535)1/2^-$, $N(1650)1/2^-$, and
$N(1710)1/2^+$ in addition to $N(1440)1/2^+$, though their data are
not much known.  In Model II, we further take into account
$N(1520)3/2^-$, $N(1675) 5/2^-$ and $N(1680)5/2^+$ together with
those included in Model I.

The present work is sketched as follows: In Section II, we explain the
general formalism for the $f_0(980)$ and $f_0(500)$
photoproductions. In Section III, we present the numerical results
separately for Model I and Model II, and discuss their physical
implications. The last Section is devoted to the summary and the
conclusion of the present work. We also discuss perspectives of future
works in the last Section.
\section{General formalism}
\begin{figure}[htp]
\centering
\includegraphics[scale=0.4]{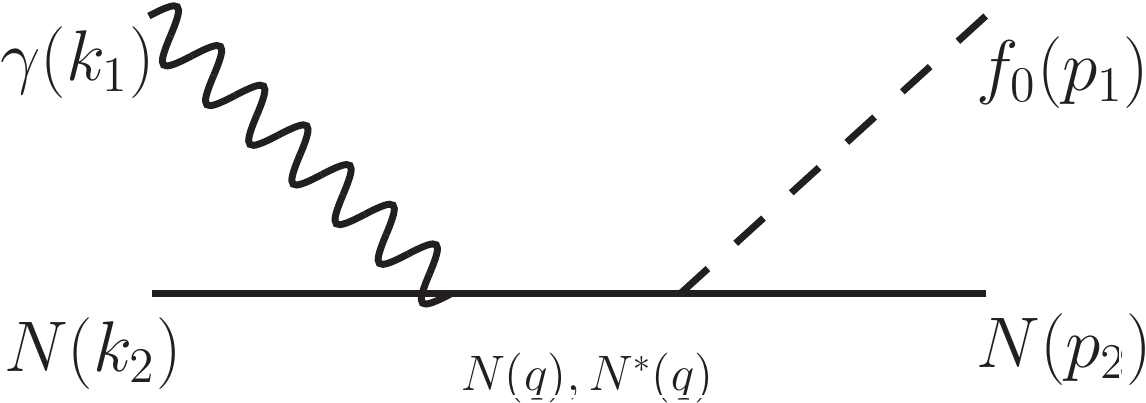} \hfill
\includegraphics[scale=0.4]{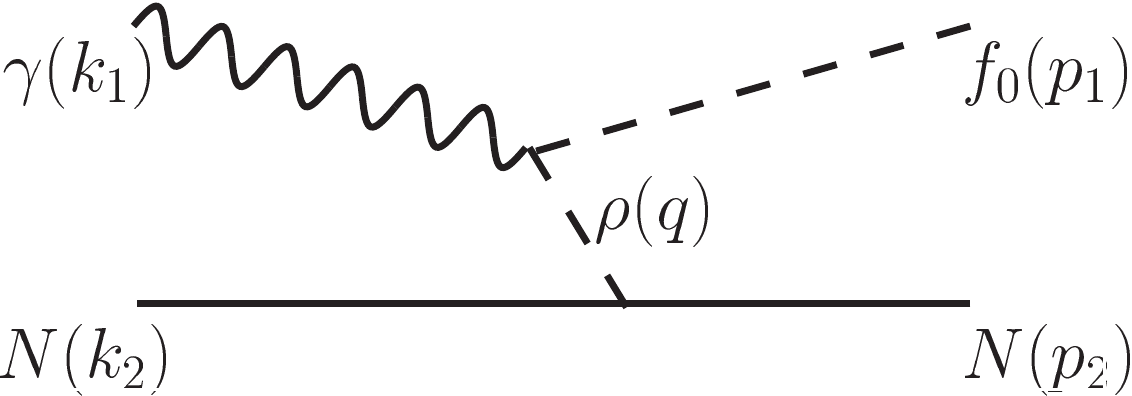} \hfill
\includegraphics[scale=0.4]{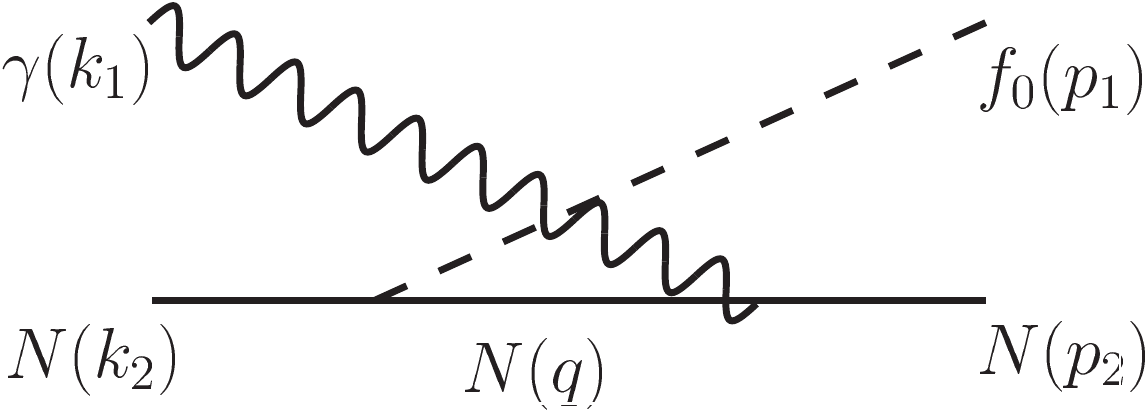}
\caption{Feynman diagrams for $f_0$ photoproduction at the tree
  level. Each diagram corresponds to the $s$ channel, the $t$ channel, and
the $u$ channel in order.}
\label{fig:1}
\end{figure}
We start with the tree-level Feynman diagrams relevant to the $\gamma p
\to f_0(980) p$ and $\gamma p\to f_0(500)p$ reactions based on the
effective Lagrangian approach, as depicted in
Fig.~\ref{fig:1}. Note that we will not consider any contribution from
$N^*$ resonances to $f_0(980)$ photoproduction, since the threshold
energy of its production is rather high. There is no $N^*$ above 1900
MeV, which can decay into $f_0(980) N$. Thus, we take only into
account  nucleon exchange both in the $s$- and $u$- channels.
As far as the $t$-channel diagram is concerned, $\rho$-meson exchange
comes into play. However, the $\gamma N\to f_0(980) N$ reaction was
experimentally measured at higher photon energies
($E_\gamma=3.0-3.8$ GeV)~\cite{Battaglieri:2008ps,
  Battaglieri:2009aa}, which is quite far beyond threshold.
Since the method of effective Lagrangians is devised for describing
the mechanism of hadronic reactions near threshold, we need to revise
it to explain $f_0(980)$ photoproduction at higher $E_\gamma$.
We will employ a hybridized Regge approach in which
the Feynman propagator in the $t$ channel is replaced with the
Regge propagator for $\rho$-meson exchange,
while the coupling constants and spin structures are
taken from the effective Lagrangians. This approach was successfully
used for describing the productions of strange and charmed
hadrons~\cite{Kim:2014qha, Kim:2015ita, Kim:2016cxr}. A virtue of
using this hybridized Regge model is that we can use the reggezied
$\rho$ meson in the $t$ channel also for $f_0(500)$ photoproduction
with the same parameters fixed in $f_0(980)$ photoproduction.

As for the $\gamma N\to f_0(500)N$ reaction,
we include the $\rho$ Reggeon in the $t$ channel with the same
parameters used in $f_0(980)$ photoproduction. In the $s$- and
$u$-channels, we consider nucleon exchange.  Since the threshold
energy of $f_0(500)$ photoproduction is about $1.4-1.5$ GeV, we
need to introduce in the $s$ channel the $N^*$ resonances that decay
only into $(\pi\pi)_{S-\mathrm{wave}}^{I=0}N$. To study the contributions of
the $N^*$ resonances, we develop two different models: In Model I, we
include $N(1440)1/2^+$, $N(1535)1/2^-$, $N(1650)1/2^-$, and
$N(1710)1/2^+$, while Model II further contains $N(1520)3/2^-$ ,
$N(1675) 5/2^-$ and $N(1680)5/2^+$with higher spins in addition to
those in Model I. Kinematics of the $\gamma p\to f_0 p$ reactions are
given as shown in Fig.~\ref{fig:1}: $k_1$ and $k_2$  stand for the
four momenta respectively for the photon and the nucleon in the
initial state, whereas $p_1$ and $p_2$ designate those respectively
for the $f_0$ and the nucleon in the final state.

\subsection{Model I}
As we have already briefly mentioned in Introduction, there are ten
excited nucleon resonances that can decay into $f_0(500) N$. However,
if we consider $N^*$ with overall status 3 or 4 star, then there are
only seven $N^*$ resonances. In order to scrutinize the
effects of the $N^*$ resonances carefully, we first introduce the
pertinent $N^*$ resonances only with spin $1/2$ to describe $f_0(500)$
photoproduction. We call this \textit{Model I}.

To compute the Feynman invariant amplitudes for $f_0$ photoproduction,
we use the following effective Lagrangians~\cite{Ryu:2012tw,
Kim:2011rm,Kim:2012pz,Nakayama:2006ty}:
\begin{align}
\mathcal{L}_{\gamma f_{0} \rho}
& =  \frac{eg_{\gamma f_{0} \rho}}{2m_{\rho}}
  \partial_{\mu}A_{\nu} (\partial^{\mu}\rho^{\nu}
- \partial^{\nu}\rho^{\mu})f_{0},\cr
\mathcal{L}_{\gamma NN}
&=-\bar{N}\left(e_N\gamma_{\mu}A^{\mu}
-\frac{e\kappa_{N}}{2m_{N}}\sigma_{\mu\nu}\partial^{\nu}A^{\mu}\right)N, \cr
\mathcal{L}_{\rho NN} & = -g_{\rho NN}\bar{N}\left(\gamma_{\mu}
\rho^{\mu}-\frac{\kappa_{\rho}}{2m_{N}}\sigma_{\mu\nu}
\partial^{\nu}\rho^{\mu}\right)N,   \cr
\mathcal{L}_{f_0 NN}&= g_{f_0 NN}f_0\bar{N}N, \cr
\mathcal{L}_{\gamma NN^*\left(\frac{1}{2}^{\pm}\right)}
&= \frac{ef_1}{2m_N} \bar{N} \Gamma^{(\mp)} \sigma_{\mu \nu}
\partial^{\nu}A^{\mu} N^* + \mathrm{H.c.},\cr
\mathcal{L}_{f_0 NN^*\left(\frac{1}{2}^{\pm}\right)}
&= \pm g_{f_0 NN^*} f_0\bar{N} \Gamma^{(\mp)}N^* + \mathrm{H.c.},
\label{eq:EffLag}
\end{align}
where $A$, $N$, $\rho$, and $f_0$ denote the photon, the nucleon, the
$\rho(770,1^-)$, and the $f_0$ fields, respectively.
The $N^*$ represents a
field for the excited nucleon. The values of the coupling constants
given in the Lagrangians will be discussed later. The matrix
$\Gamma^{(\pm)}$ depends on the parity of the $N^*$ resonance and is
defined as
\begin{align}
\Gamma^{(+)} = \gamma_5,\;\;\;\; \Gamma^{(-)} = \bm{1}.
\end{align}

Based on the effective Lagrangians in Eq.~(\ref{eq:EffLag}),
we can compute the Feynman invariant amplitudes for each channel as
follows:
\begin{align}
-i\mathcal{M}_{t\left(\rho\right)}
=&\frac{ieg_{\gamma f_0 \rho}g_{\rho NN}}{2m_{\rho}}\bar{u}(p_2)
 \frac{1}{t-m_\rho^2}
 \Bigl[ \epsilon^\alpha (k_1 \cdot p_1) - k_1^\alpha
 (\epsilon \cdot p_1) \Bigr]
 \Bigl[ \gamma_\alpha - \frac{i\kappa_\rho}{2m_N} \sigma_{\alpha\beta} q_t^\beta
 \Bigr] u(k_2),
\nonumber   \\
-i\mathcal{M}_{s\left(N\right)}
=& i g_{f_0 NN} \bar{u}(p_2) \frac{\slashed{q_s}+m_N}{s-m_N^2}
   \slashed{\epsilon}
   \left[e_{N}-\frac{e\kappa_{N}}{2m_{N}}\slashed{k_1}\right] u(k_2),
\nonumber   \\
-i\mathcal{M}_{u\left(N\right)}
=& i g_{f_0 NN} \bar{u}(p_2) \slashed{\epsilon}
   \left[e_{N}-\frac{e\kappa_{N}}{2m_{N}} \slashed{k_1} \right]
   \frac{\slashed{q_u}+m_N}{u-m_N^2} u(k_2),
\label{Am_u}
\end{align}
where the momentum transfers are given as $q_t = p_1 - k_1$,
$q_s = k_1 + k_2$, and $q_u = p_2 - k_1 $.
$e_N$ is the electric charge of the nucleon and $\epsilon_\mu$
indicates the polarization vector of the photon.
The $u(p_2)$ and $u(k_2)$ denote the Dirac spinors of the
outgoing nucleon and the incoming nucleon, respectively.
The $m_N$ and $m_{\rho}$ stand for the masses of the nucleon and
$\rho$ meson, respectively. $t$, $s$, and $u$ represent the Mandelstam
variables and are defined as
\begin{equation}
t  = (p_1-k_1)^2,\;\;\;\;s = (k_1+k_2)^2,\;\;\;\;u = (p_2-k_1)^2.
\end{equation}

The effective Lagrangian approach has been successfully used to
describe hadronic reaction in the low-energy region. However, when the
photon energy increases, the results from ths effective
Lagrangians start to deviate from the data and do not even satisfy the
unitarity~\cite{Kim:2015ita}. Since the CLAS data on $f_0(980)$
photoproduction was conducted at $E_\gamma=3.0$ GeV and $E_\gamma
=3.8$ GeV, which is far from threshold, the effective Lagrangian
method  is not suitable to explain the data. Thus, we employ a
hybridized Regge model as already mentioned previously. In this
approach, the Reggeon in the $t$-channel is governed by the Regge
trajectory of the $\rho$ meson, which is well known
already~\cite{Schmid:1968zz, Shapiro:1969km, Jackson:1970aw,
  Donnachie:2002, Sertorio:1969ud, Guidal:1997hy}.
We replace the Feynman propagator with the Regge one
$\mathcal{P}_{\rho}^{\mathrm{R}}$ in the $t$
channel~\cite{Donnachie:2002,Sertorio:1969ud,Guidal:1997hy}
\begin{equation}
\frac{1}{t-m_{\rho}^2}  \Rightarrow \mathcal{P}_{\rho}^{\mathrm{R}}(s,t),
\end{equation}
where the Regge propagator is defined as
\begin{align}
\mathcal{P}_{\rho}^{\mathrm{R}}(s,t) =
\left( \frac{s}{s_{\rho}} \right)^{\alpha_{\rho}(t)-1}
\frac{1}{\sin[\pi \alpha_\rho(t)]}
\frac{\pi \alpha'_\rho}{\Gamma[\alpha_\rho(t)]}.
\label{eq:ReggePropagator}
\end{align}
Here, $\alpha_\rho(t)$ denotes the Regge trajectory for the $\rho$
meson. $s_\rho$ indicates the energy scale parameter for the
corresponding Reggeon and is set to be equal to
$1.0\,\mathrm{GeV}^2$. The Regge trajectory for the
$t$-channel is taken from Ref.~\cite{Guidal:1997hy}
\begin{align}
\alpha_{\rho}(t)&=0.55 +0.8t.
\end{align}

We adopt a degenerate propagator~\cite{Guidal:1997hy,Laget:2005be}
with a constant phase ($1$). A rotating phase ($e^{-i\pi\alpha(t)}$)
is also acceptable. Although the nondegenerate phase
((1-exp[-i$\pi\alpha(t)$])/2) is general for the $\rho$
trajectory~\cite{Colllins:1984}, 
it is excluded from our consideration
because the corresponding result for the $d\sigma/dt$ reveals a dip
structure near $-t \simeq 0.7\,\mathrm{GeV^2}$ in the range $E_\gamma$ = 
(3.0 - 3.8) GeV which definitely deviates from the expemental
data~\cite{Battaglieri:2008ps}.

In addition, we introduce the scaling factor for the $\rho$ meson
Reggeon exchange
\begin{align}
  \label{eq:scaling}
C(t) = \frac{a_\rho}{\left(1-t/\Lambda_{\rho}^2\right)^2},
\end{align}
which are often included to explain experimental values of the cross
sections. However, we find that the results are not sensitive to the
parameters $a_\rho$ and $\Lambda_\rho$ of the scaling factor.

\begin{table}[htp]
\centering
\begin{tabular}{ c | c c c  c}
\hline\hline
Particle &  $J^P$ & Mass [GeV] & Width [GeV] & Status\\
\hline
$\textrm{N}(1440)$ & $\frac{1}{2}^+$ & 1.440 & 0.300 & **** \\
$\textrm{N}(1535)$ & $\frac{1}{2}^-$ & 1.535 & 0.150 & **** \\
$\textrm{N}(1650)$ & $\frac{1}{2}^-$ & 1.655 & 0.150 & **** \\
$\textrm{N}(1710)$ & $\frac{1}{2}^+$ & 1.710 & 0.100 & ***  \\
\hline \hline
\end{tabular}
\caption{Spin 1/2 resonances~\cite{PDG2016} for
$\gamma N \to f_0(500) N$ photoproduction in Model 1.}
\label{tab:1}
\end{table}
While $f_0(980)$ photoproduction does not require any contribution
from the $N^*$ resonances because of its high threshold energy, we
need to consider them for the explanation of the $\gamma p \to
f_0(500) p$ reaction. As mentioned earlier, we will take into account
the spin $1/2$ resonances in Model I: $N(1440)1/2^+$, $N(1535)1/2^-$,
$N(1650)1/2^-$, and $N(1710)1/2^+$ as listed in Table~\ref{tab:1}. Since
all of them have rather large widths, we will include the finite width
in each propagator for the $N^*$. Then, the Feynman amplitudes are
derived as
\begin{align}
-i \mathcal{M}_{s\left(N^*\left(\frac{1}{2}^+\right)\right)}
=& \frac{ef_1g_{f_0NN^*}}{2m_N}\bar{u}(p_2)
 \frac{i(\slashed{q_s} +
   m_{N^*})}{s-(m_{N^*}-\frac{i}{2}\Gamma_{N^*})^2}
 \slashed{k_1}\slashed{\epsilon}u(k_2),
\nonumber   \\
-i \mathcal{M}_{s\left(N^*\left(\frac{1}{2}^-\right)\right)}
=& \frac{ef_1g_{f_0NN^*}}{2m_N}\bar{u}(p_2)
 \frac{i(-\slashed{q_s} +
   m_{N^*})}{s-(m_{N^*}-\frac{i}{2}\Gamma_{N^*})^2}
 \slashed{k_1}\slashed{\epsilon}u(k_2),
\label{eq:amp}
\end{align}
where the first term is the amplitude for the $N^*$ with positive parity
whereas the second one represents that for the negative-parity $N^*$.
The coupling constants $f_1$ and $g_{f_0NN^*}$ denote respectively the
generic photocouplings and strong coupling constants for the
corresponding $N^*$ resonances. $m_{N^*}$ and $\Gamma_{N^*}$ represent
the corresponding masses and the decay widths of the $N^*$,
respectively. Note that, for the propagators in Eq.(\ref{eq:amp}), we consider the pole positions of the $N^*$ resonances in
the complex plane.

\subsection{Model II}
In Model II, we additionally consider the $N^*$ resonances with
spins $3/2$ and $5/2$ in the $s$-channel. We include the
$N(1520)3/2^-$, $N(1675)5/2^-$, and $N(1680)5/2^+$ in the $s$ channel
as listed in Table~\ref{t2}~\cite{PDG2016}.
\begin{table}[htp]
\centering
\begin{tabular}{ c | c c c  c}
\hline\hline
Particle &  $J^P$ & Mass [GeV] & Width [GeV] & Status\\
\hline
$\textrm{N}(1520)$ & $\frac{3}{2}^-$ & 1.525 & 0.115 & **** \\
$\textrm{N}(1675)$ & $\frac{5}{2}^-$ & 1.675 & 0.150 & **** \\
$\textrm{N}(1680)$ & $\frac{5}{2}^+$ & 1.685 & 0.130 & **** \\
\hline \hline
\end{tabular}
\caption{The excited nucleon resonances~\cite{PDG2016} for
$f_0(500)$ photoproduction in Model II}
\label{t2}
\end{table}
The relevant effective Lagrangians for the $N^*$ resonances are given
as follows
\begin{align}
\mathcal{L}_{\gamma NN^*\left(\frac{3}{2}^-\right)}
&= \frac{ief_1}{2m_N}\bar{N}^*_{\mu} F^{\mu \nu} \gamma_{\nu}N
-\frac{ef_2}{(2m_N)^2}\bar{N}^*_{\mu}F^{\mu \nu} \partial_{\nu}N
 + \mathrm{H.c.}, \nonumber \\
\mathcal{L}_{\gamma NN^*\left(\frac{5}{2}^{\pm}\right)}
&= \pm \frac{ef_1}{(2m_N)^2}\bar{N}^*_{\mu \alpha} \partial^{\alpha}
F^{\mu \nu}\Gamma_{\nu}^{(\mp)}N
\pm \frac{ief_2}{(2m_N)^3}\bar{N}^*_{\mu \alpha}\partial^{\alpha}
F^{\mu \nu}\Gamma^{(\mp)}\partial_{\nu}N
 + \mathrm{H.c.}, \nonumber \\
\mathcal{L}_{f_0  NN^*\left(\frac{3}{2}^-\right)}
&= \frac{g_{f_0 NN^*}}{m_{f0}}(\partial_{\mu}f_0)\bar{N}N^{* \mu}
 + \mathrm{H.c.}, \nonumber \\
\mathcal{L}_{f_0NN^*\left(\frac{5}{2}^{\pm}\right)}
&= i \frac{g_{f_0 NN^*}}{m_{f_0}^2}\bar{N}(\partial_{\mu}\partial_{\nu}f_0)
\Gamma^{(\mp)}N^{*\mu\nu} + \mathrm{H.c.},
\end{align}
where $f_1$, $f_2$, and $g_{f_0NN^*}$ denote the photocouplings and
the strong coupling constants, respectively. They can be determined by
using the experimental data on the photon decay amplitudes and the
decay widths $\Gamma_{N^*\to f_0(500)N}$. Then, the Feynman invariant
amplitudes for the $s$-channel are derived as
\begin{align}
-i \mathcal{M}_{s\left(N(1520)\right)}
=&\; i p_{1}^{\mu} \bar{u}(p_2) \Delta_{\mu \alpha}(q_s, m_{N^*})
(k_1^{\alpha}\epsilon^{\beta}-k_1^{\beta}\epsilon^{\alpha}) \cr
&\times \left(  \frac{ef_1g_{f_0NN^*}}{2m_Nm_{f_0}}\gamma_{\beta} +
  \frac{ef_2g_{f_0 NN^*}} {4m_N^2m_{f_0}}k_{2 \beta} \right)u(k_2),
\nonumber \\
-i \mathcal{M}_{s\left( N(1675)\right)}
=&\; \bar{u}(p_2)p_{1}^{\rho}p_{1}^{\sigma}\gamma_5 \Delta_{\rho
   \sigma;\mu \alpha}(q_s, m_{N^*})
k_{1}^{\alpha}(k_1^{\mu}\epsilon^{\nu}-k_1^{\nu}\epsilon^{\mu}) \cr
& \times \biggl(\frac{ief_1g_{f_0NN^*}}{4m_N^2m_{f_0}^2}\gamma_{\nu}
+\frac{ief_2g_{f_0NN^*}}{8m_N^3m_{f_0}^2}k_{2\nu}\biggl)\gamma_5u(k_2),
\nonumber \\
-i \mathcal{M}_{s\left( N(1680)\right)}
=&\; -\bar{u}(p_2)p_{1}^{\rho}p_{1}^{\sigma}
\Delta_{\rho \sigma;\mu \alpha}(q_s, m_{N^*})
k_{1}^{\alpha}(k_1^{\mu}\epsilon^{\nu}-k_1^{\nu}\epsilon^{\mu})  \cr
& \times \biggl(\frac{ief_1g_{f_0NN^*}}{4m_N^2m_{f_0}^2}\gamma_{\nu}
+\frac{ief_2g_{f_0NN^*}}{8m_N^3m_{f_0}^2}k_{2\nu}\biggl)u(k_2),
\end{align}
where $\Delta_{\mu \alpha}$ and $\Delta_{\rho \sigma;\mu \alpha}$ indicate the
Rarita-Schwinger propagators for the $N^*$ resonances with spin $3/2$ and
$5/2$, respectively, defined
as~\cite{Behrends:1957rup,Rushbrooke:1966zz,Chang:1967zzc,Oh:2011}
\begin{align}
\Delta_{\mu \alpha}(q,m_{N^*})
&=\frac{i (\slashed{q} + m_{N^*})}
  {s-\left(m_{N^*} - \frac{i}{2}\Gamma_{N^*} \right)^2}
  S_{\mu \alpha}(q, m_{N^*}), \cr
\Delta_{\alpha \beta; \mu \nu}(q,m_{N^*})
&=\frac{i (\slashed{q} + m_{N^*})}
  {s-\left(m_{N^*}- \frac{i}{2}\Gamma_{N^*} \right)^2}
  S_{\alpha \beta;\mu \nu}(q, m_{N^*}).
\end{align}
Here, $S_{\mu \alpha}$, $\bar{g}_{\mu \alpha}$, and
$\bar{\gamma}_{\mu}$ are expressed as
\begin{align}
S_{\mu\alpha}(q, m_{N^*})
&= -\bar{g}_{\mu\alpha}
+\frac{1}{3}\bar{\gamma}_{\mu}\bar{\gamma}_{\alpha}, \cr
S_{\alpha\beta;\mu\nu}(q,m_{N^*})
&= \frac{1}{2}\left(\bar{g}_{\alpha\mu}\bar{g}_{\beta\nu}
+\bar{g}_{\alpha\nu}\bar{g}_{\beta\mu}\right)
-\frac{1}{5}\bar{g}_{\alpha\beta}\bar{g}_{\mu\nu}  \cr
& -\frac{1}{10}\left(\bar{\gamma}_{\alpha}\bar{\gamma}_{\mu}\bar{g}_{\beta\nu}
+\bar{\gamma}_{\alpha}\bar{\gamma}_{\nu}\bar{g}_{\beta\mu}
+\bar{\gamma}_{\beta}\bar{\gamma}_{\mu}\bar{g}_{\alpha\nu}
+\bar{\gamma}_{\beta}\bar{\gamma}_{\nu}\bar{g}_{\alpha\mu}\right),   \cr
\bar{g}_{\mu \alpha}
&= g_{\mu\alpha}-\frac{q_\mu q_\alpha}{m_{N^*}^2},\,
\bar{\gamma}_{\mu}
= \gamma_{\mu} - \frac{q_\mu\slashed{q}}{m_{N^*}^2}.
\end{align}
\subsection{Parameters and form factors}
The coupling constant for the $\rho NN$ vertex is the most important
parameter to describe $f_0(980)$ photoproduction, since the
$t$-channel governs the production mechanism of the $\gamma N \to
f_0(980)N$ reaction. The coupling constant $g_{\rho NN}$ and
$\kappa_{\rho}$ are well known from $NN$ potentials. For example,
$g_{\rho NN}= 3.25$ was used in the full Bonn
potential~\cite{Machleidt:1987hj}, while the Nijmegen group employed
$g_{\rho NN}= 2.76$~\cite{Rijken:2006en}.
On the other hand, smaller coupling constants
$g_{\rho NN}=2.6$ and $\kappa_{\rho}=3.7$ have been exploited in Regge
models for photoproduction of the pion and of the charged $\rho^{\pm}$
meson~\cite{Yu:2011zu, Yu:2016psj}.
We use $g_{\rho NN}=3.25$ and the ratio of the vector and
tensor couplings $\kappa_{\rho}=6.1$~\cite{Machleidt:1987hj}.

The $f_0(980)NN$ coupling constants is taken from
Ref.~\cite{Schumacher:2007ez}: $g_{f_0(980) NN}= 5.8$. The
$f_0(500) NN$ coupling constant $g_{f_0(500)NN}$ is a crucial one
which explains the mid-range strong attraction in the $NN$
interaction. Its value is given in a wide numerical range. For
example, the full Bonn potential suggests $g_{\sigma' NN}=
8.46$~\cite{Machleidt:1987hj} with $m_{\sigma'}=550$ MeV.
The notation $\sigma'$ in Ref. ~\cite{Machleidt:1987hj}
was introduced to emphasize the fact that $\sigma'$ represents only an
effective description of correlated $2\pi$ exchange in $S$ wave. In
the one-boson-exchange (OBE) Bonn NN potential, two different
$\sigma$s were introduced, i.e. $\sigma_1$ in the isovector ($T=1$)
channel and $\sigma_2$ in the isoscalar channel ($T=0$). The
charge-dependent OBE Bonn $NN$ potential has even several different
values for different partial waves in $NN$
scattering~\cite{Machleidt:2000ge}. One has to keep in mind that the
coupling constant for the $\sigma NN$ vertex describes effectively
whole $2\pi$ and even $\pi\rho$ exchanges. On the other hand, the
Nijmegen  soft-core (NSC) potential includes two different scalar
mesons, flavor SU(3) symmetry and ideal mixing being
used~\cite{Rijken:1998yy}. Depending on whether the scalar mesons
constitute quark-antiquark pairs or tetraquarks,
Ref.~\cite{Rijken:1998yy} suggested two different values for the
$\sigma NN$ and $f_0(980) NN$ coupling constants with different flavor
content. Since the $f_0(500)$ in the present work corresponds to the
$S$-wave correlated $2\pi$, we will take the value of $g_{\sigma' NN}$
from the full Bonn potential, i.e. $g_{f_0(500) NN}=8.46$. This
choice is reasonable, since $f_0(500)$ in the present work indeed
corresponds to the resonance in the $S$-wave correlated $2\pi$
channel. Note that $\sigma'$-exchange in the full Bonn potential was
later replaced by the explicit $S$-wave correlated
$2\pi$-exchange~\cite{Kim:1994ce,Reuber:1995vc}. 
Of course, there is one caveat: Since the $f_0(500)$ meson
has a very broad width, a single coupling constant is a rather crude
approximation. A more complete work considering $f_0(500)$ as
$S$-wave correlated $2\pi$ resonance will be considered in a future
work. 

The mass of the $f_0(500)$ meson brings out another ambiguity in dealing
with $f_0(500)$ photoproduction. Because of the broad width of
$f_0(500)$ ($\Gamma=(400-700)$ MeV), it is rather difficult to
determine its mass exactly.  The Particle Data Group estimated the
pole mass of $f_0(500)$ to be $(400-500)-i(200-350)$
MeV~\cite{PDG2016}. It implies that we need to examine
carefully the $f_0(500)$ mass dependence of observables. If it is
taken to be larger than $m_{f_0(500)}=500$ MeV, the threshold energy
can be larger than the mass of the $N(1440)$. Thus, the total cross
section of the $\gamma N\to f_0(500) N$ could reveal different
behavior from that with the lower value of the $f_0(500)$ mass. We
will discuss in detail the dependence of the total cross section on
$m_{f_0(500)}$ in the next Section.

In order to determine the strong coupling constants for the excited
baryons, we assume that  the $f_0(500)$ meson consists mainly of the
correlated $2\pi$ state in $S$-wave. Then, we can use the decay modes
of $N^*\to (\pi^+\pi^-)_{S-\mathrm{wave}}^{I=0}$ to fix the
corresponding coupling constants by using the partial decay widths
defined as
\begin{align}
\Gamma(N^* {\textstyle(\frac{1}{2}^{\pm})} \rightarrow f_0 N )
&=  \frac{g_{f_0 NN^*}^2 }{4\pi}\frac{|\bm{p}|}{m_{N^*}}\left(E_N \pm m_N\right), \cr
\Gamma(N^*{\textstyle(\frac{3}{2}^{\pm})}\rightarrow f_0 N )
&= \frac{1}{3} \frac{g_{f_0 NN^*}^2}{4\pi}\frac{|\bm{p}|^3}{m_{N^*} m_{f_0}^2}
  \left(E_N \mp m_N\right), \cr
\Gamma(N^* {\textstyle(\frac{5}{2}^{\pm})}\rightarrow f_0 N )
&=  \frac{2}{15}\frac{g_{f_0 NN^*}^2}{4\pi}\frac{|\bm{p}|^5}{m_{N^*} m_{f_0}^4}
  \left(E_N\pm m_N\right),
\label{eq:pdw_Nstr}
\end{align}
where the partial decay width, $\Gamma(N^*\to f_0 N)$ is given as
$\Gamma_{N^*}^{\mathrm{BW}} \times \mathrm{Br}(N^* \to f_0
N)$. $\Gamma_{N^*}^{\mathrm{BW}}$ denotes the Breit-Wigner total decay
width and $\mathrm{Br}(N^* \to f_0N)$ the branching ratio.
$|\bm{p}|$ represents the magnitude of the final-state momentum
defined as
\begin{align}
|\mathbf{p}| &= \frac{1}{2m_{N^*}}
\sqrt{[m_{N^*}^2 - (m_N+m_{f_0})^2][m_{N^*}^2 - (m_N-m_{f_0})^2]}.
\end{align}
The results of the strong coupling constants are shown in
Table. \ref{tab:3}
\begin{table}[htp]
\begin{tabular}{cccc|ccc}
\hline\hline
$g_{f_0NN(1440)}$ & $g_{f_0NN(1535)}$ &$g_{f_0NN(1650)}$
  &$g_{f_0NN(1710)}$ & $g_{f_0NN(1520)}$ & $g_{f_0NN(1675)}$
  &$g_{f_0NN(1680)}$   \\  \hline
$\pm$3.88 & $\pm$2.55 & $\pm$0.96 & $\pm$0.15 &
  $\pm$0.85  &$\pm$9.85 & $\pm$2.29 \\
\hline\hline
\end{tabular}
\caption{The strong coupling constants for the $N^*$ resonances. The
  first four coupling constants correspond to those for the excited
  nucleons with spin $1/2$, whereas the next three ones are those for
  the $N^*$ resonances with spin is $3/2$ and $5/2$.}
\label{tab:3}
\end{table}

Concerning the photocoupling constants for $\rho$-meson exchange, we
use $g_{\gamma f_0(500) \rho}= 0.25$~\cite{Achasov:2002jv} from the
measurement of the SND Collaboration, and $g_{\gamma f_0(980) \rho}
\approx 0.21$~\cite{Kalashnikova:2005zz,Branz:2007xp} from the molecular
$K \bar K$ model for $\Gamma_{f_0 \to \rho \gamma}$~\cite{Obukhovsky:2009th}.
On the other hand, we can utilize the helicity amplitudes given in the
Particle Data Group~\cite{PDG2016} to find the photocoupling
constants $f_1$ and $f_2$ for excited nucleons. Since the helicity
amplitude are expressed as~\cite{Oh:2007jd}
\begin{align}
A_{1/2}\left(\frac{1}{2}^{\pm}\right)&= \mp \frac{ef_1}{2m_N}
\sqrt{\frac{k_{\gamma}m_{N^*}}{m_N}},  \cr
A_{1/2}\left(\frac{3}{2}^{\pm}\right)&= \mp\frac{e\sqrt{6}}{12}
\sqrt{\frac{k_{\gamma}}{m_Nm_{N^*}}}
\left[f_1+\frac{f_2}{4m_N^2}m_{N^*}(m_{N^*}\mp m_N)\right], \cr
A_{3/2}\left(\frac{3}{2}^{\pm}\right)&=\mp\frac{e\sqrt{2}}{4m_N}
\sqrt{\frac{k_{\gamma}m_{N^*}}{m_N}}
\left[f_1\mp\frac{f_2}{4m_N}(m_{N^*}\mp m_N)\right], \cr
A_{1/2}\left(\frac{5}{2}^\pm\right)&= \pm\frac{e}{4\sqrt{10}}
\frac{k_{\gamma}}{m_N} \sqrt{\frac{k_{\gamma}}{m_Nm_{N^*}}}
\left[f_1+\frac{f_2}{4m_N^2}m_{N^*}(m_{N^*}\pm m_N)\right], \cr
A_{3/2}\left(\frac{5}{2}^\pm\right)&= \pm\frac{e}{4\sqrt{5}}
\frac{k_{\gamma}}{m_N^2} \sqrt{\frac{k_{\gamma}m_{N^*}}{m_N}}
\left[f_1\pm\frac{f_2}{4m_N}(m_{N^*}\pm  m_N)\right],
\end{align}
where $k_{\gamma}$ stands for the photon decay momentum in the center
of mass (CM) frame and is expressed as
\begin{align}\label{}
k_{\gamma}=\frac{m_{N^*}^2-m_N^2}{2m_{N^*}},
\end{align}
we obtain the photocouplings for the electromagnetic transitions
$N^*\to \gamma N$ as listed in Table~\ref{tab:4}.
\begin{table}[htp]
\begin{tabular}{cccc}
\hline\hline
$f_{\gamma NN(1440)}$ & $f_{\gamma NN(1535)}$ & $f_{\gamma NN(1650)}$
  & $f_{\gamma NN(1710)}$ \\
0.47 & 0.81 & 0.28 & -0.24 \\
\hline
$f_{1\gamma NN(1520)}$ & $f_{2\gamma NN(1520)}$  & \\
4.63 & -4.92 &  \\
\hline
$f_{1\gamma NN(1675)}$ & $f_{2\gamma NN(1675)}$  & $f_{1\gamma NN(1680)}$
& $f_{2\gamma NN(1680)}$ \\
-1.34 & -2.28 & 15.24 & -13.47 \\
\hline\hline
\end{tabular}
\caption{The photon coupling constants for the $N^*$ resonances whose
  spin is 1/2 are listed in the first row.
 In the second  and the third row the photon coupling constants for
 the $N^*$ resonances whose spin is 3/2 and 5/2, respectively, are
 listed.}
\label{tab:4}
\end{table}

A hadron has a spatial size, which can be characterized by the
phenomenological form factors. Hence, one has to introduce them at
each baryon-baryon-meson vertex.
Note that each amplitude of $N$ exchanges does not satisfy the gauge
invariance in the $s$- and $u$-channels, but the sum does.
To restore the gauge invariance, we modify the Born scattering amplitudes
as
\begin{equation}
{\cal M}_{\mathrm{Born}} =
{\cal M}_t^{\mathrm{Regge}} + ( {\cal M}_s + {\cal M}_u ) F_c^2(s,\,u),
\end{equation}
where a common form factor is introduced as~\cite{Davidson:2001rk}
\begin{equation}
F_c(s,\,u) = F(s) + F(u) - F(s)F(u),
\end{equation}
with
\begin{align}
F(q^2)&=\frac{\Lambda^4}{\Lambda^4+(q^2-m_{ex}^2)^2}.
\label{eq:ff}
\end{align}
Here, $q$ indicates the off-shell four-momentum for an exchanged
hadron, $m_{ex}$ stands for its mass, and $\Lambda$ denotes a cutoff
parameter. As for the form factors in $N^*$ exchange, we also use
Eq.~(\ref{eq:ff}). However, it is not sufficient to preserve the
unitarity for the $N^*$ resonances with spin $3/2$ and $5/2$, since the
corresponding amplitudes show much stronger $q^2$ dependence so that
they are divergent as $E_\gamma$ increases. Thus, in order to tame the
divergence, we will employ a Gaussian type of the form factor for both
$N(1520)3/2^-$, $N(1675)5/2^-$ and $N(1680)5/2^+$-exchange in the
$s$-channel, defined as
\begin{align}
F(q^2) = \exp\left[ -\frac{(q^2-m_{N^*}^2)^2}{\Lambda^4}\right].
\end{align}

\section{Results and discussion}
We are now in a position to present the numerical results of our work. 
Since there exists the experimental data on the differential cross section 
of the $\gamma N\to f_0(980) N$ reaction~\cite{Battaglieri:2008ps}, we 
start with $f_0(980)$ photoproduction.

\subsection{$\gamma N \to f_0(980)N$}
In order to describe the experimental data on $d\sigma/dt$ of the
$\gamma N\to f_0(980) N$ process~\cite{Battaglieri:2008ps}, we fix the
scaling factor in Eq.~(\ref{eq:scaling}) to be $a_{\rho}=9.0$. As is
well known, while the Regge approach describes the energy dependence
very well, it cannot determine the absolute magnitude of the cross
sections. Thus, it is inevitable to introduce the scale parameter
$a_{\rho}$ to fit the cross sections. The value of the cutoff
parameter $\Lambda_{\rho}$ is selected to be 1 GeV that is a
typical order of the cut-off value. We do not fit it to avoid
additional ambiguity. The cutoff parameters in $N$- and $N^*$-exchanges
are chosen as $\Lambda_N = 0.8 \;\mathrm{GeV}$ and $\Lambda_{N^*} =
1.0 \;\mathrm{GeV}$. The values of $\Lambda_{\rho,N^*}$ are also
typical ones. Note that the cutoff value for the backward
scattering amplitude is usually smaller than that for the forward
scattering one in the literature~\cite{Kim:2011rm,Kim:2012pz}. 
Furthermore, the experimental data for the backward 
region are currently not available. So, we simply choose these values
without any fitting procedure.  

\begin{figure}[htp]
\begin{center}
\includegraphics[scale=0.6]{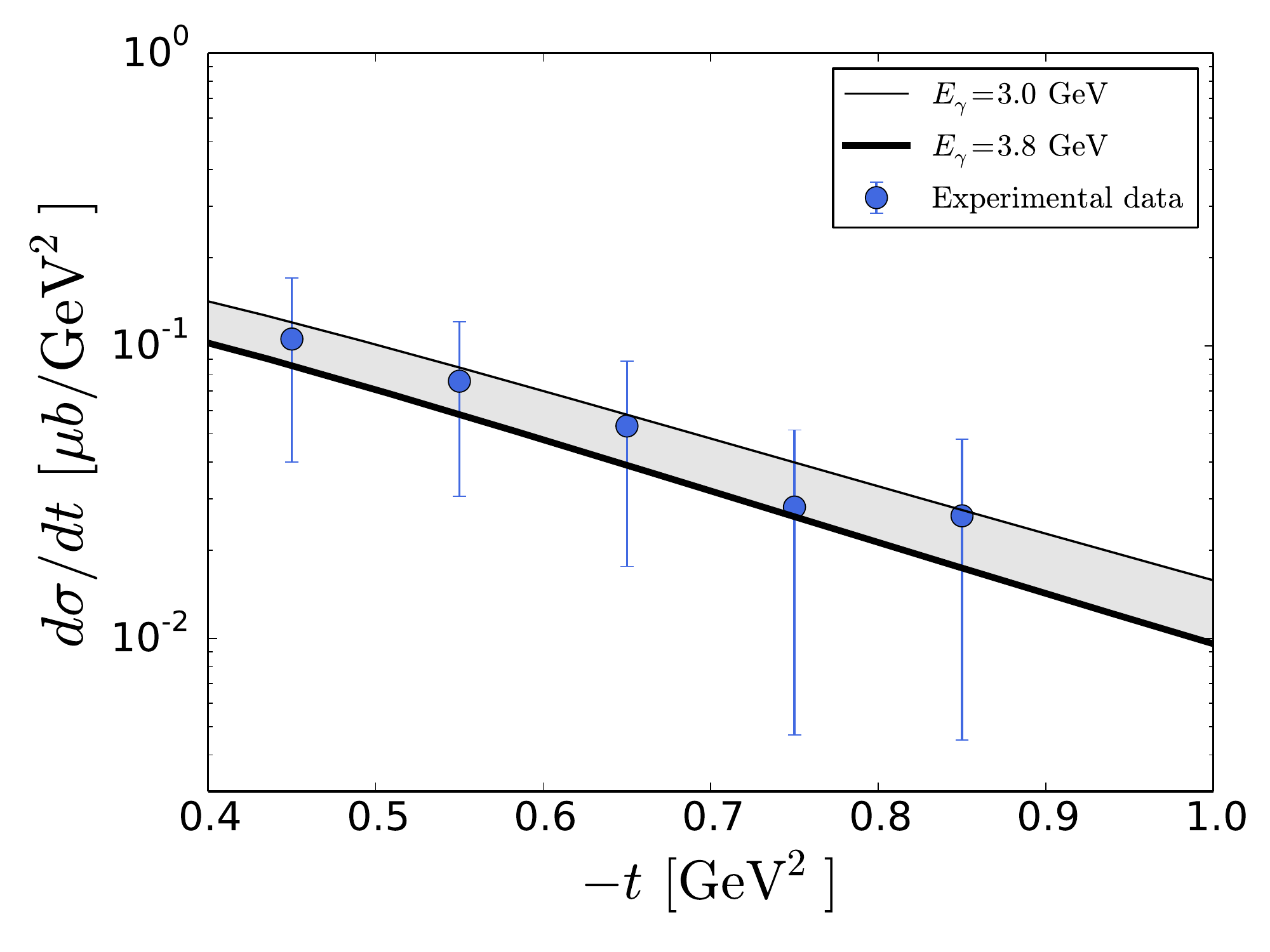}
\caption{Differential cross section of the $\gamma N \to f_0(980)N$
  reaction as a function $t$ in the range of $E_{\gamma}=(3.0-3.8) \;
  \mathrm{GeV}$. The shaded band represents the presen result. The
  experimental data are taken from Ref. \cite{Battaglieri:2008ps}.}
\label{fig:2}
\end{center}
\end{figure}
Figure~\ref{fig:2} draws the differential cross section $d\sigma/dt$
of the $\gamma N \to f_0(980)N$ reaction. The experimental data are
taken from Ref.~\cite{Battaglieri:2008ps}, where $d\sigma/dt$ were
measured within the range of the photon energy $E_{\gamma} =
(3.0-3.8)$ GeV. In order to compare the present results with the data,
we present the results as the shaded band of which the width
represents the corresponding region of $E_\gamma$. Considering the
large experimental uncertainty, the results describe the data very
well. Note that the $t$-dependence is governed by $\rho$-Reggeon
exchange in the $t$ channel.

\begin{figure}[htp]
\begin{center}
\includegraphics[scale=0.6]{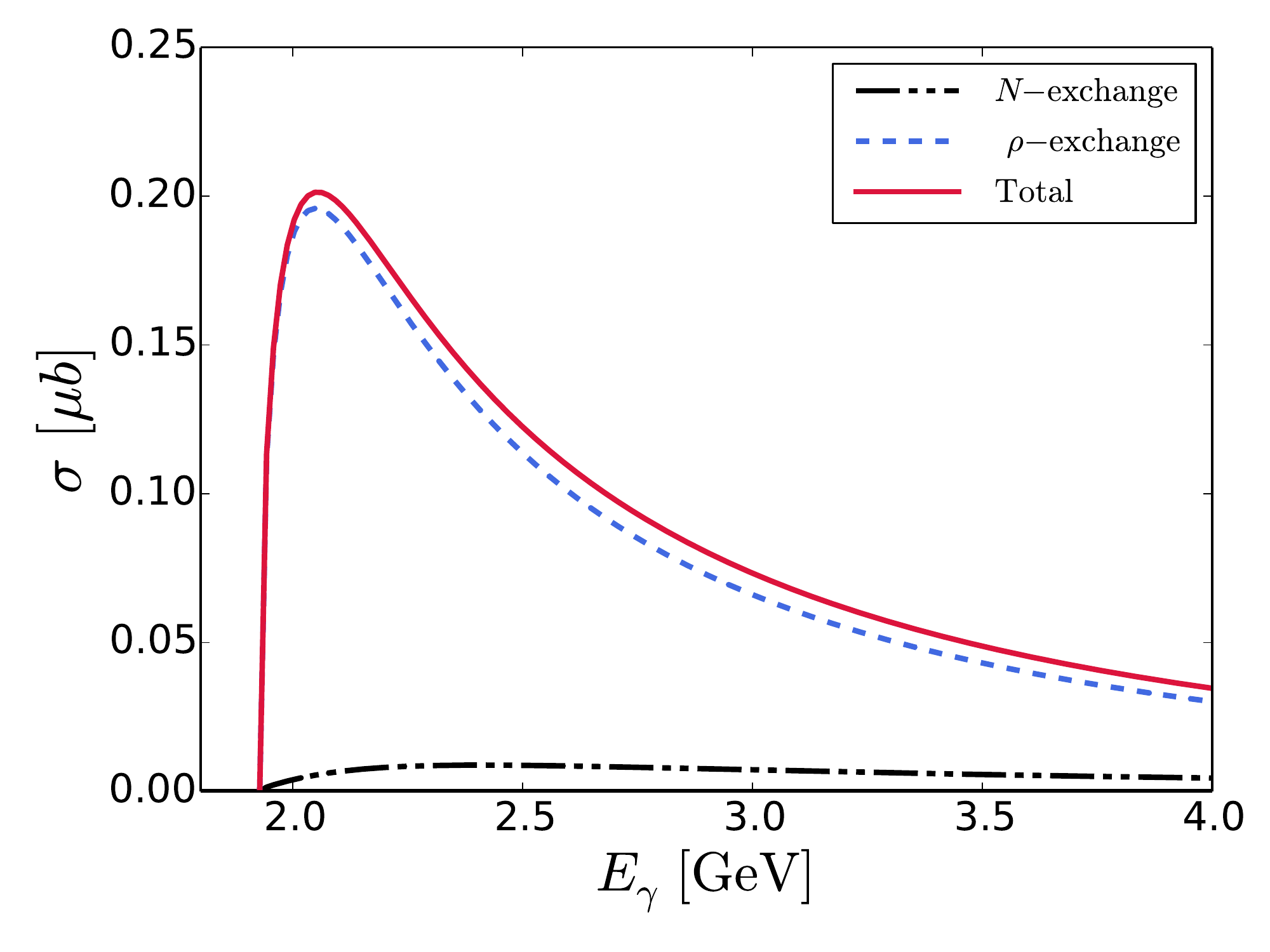}
\caption{Total cross section of the $\gamma N \to f_0(980)N$ reaction.
The dashed curve depicts the contribution of $\rho$-Reggeon exchange
in the $t$ channel, whereas the dot-dashed and double-dot-dashed ones
draw those  of $N$-Reggeon exchange and of $N$ exchange in the $u$
and $s$ channels, respectively. The solid curve represents the total
contribution.}
\label{fig:3}
\end{center}
\end{figure}
In Fig.~\ref{fig:3}, we depict the contribution of each channel to the
total cross section of the $\gamma N\to f_0(980) N$ reaction.
The $\rho$-Reggeon exchange dominates over the $s$- and
$u$-channel diagrams, whereas the $N$ exchanges have small effects
in the whole energy region.

\subsection{$\gamma N \to f_0(500)N$}
The parameters for $f_0(500)$ photoproduction in the
$t$, $s$, and $u$ channels are kept to
be the same as those in the case of the $\gamma N\to f_0(980) N$
reaction. However, the $N^*$ resonances play essential roles in
describing the $\gamma N\to f_0(500) N$ reaction in particular in the
vicinity of threshold.

\begin{figure}[htp]
\centering
\subfigure{\includegraphics[scale=0.42]{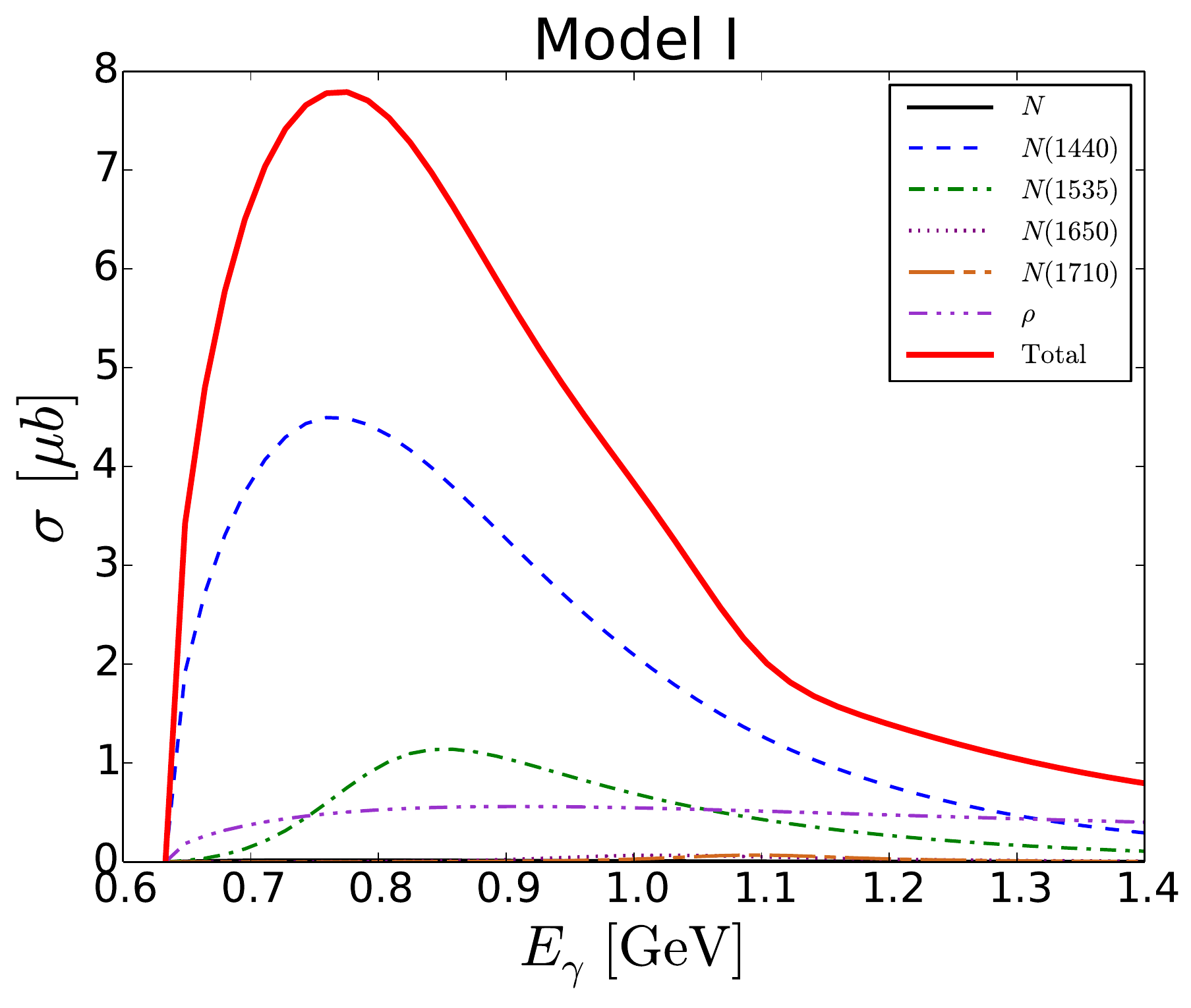}}\hfill
\subfigure{\includegraphics[scale=0.42]{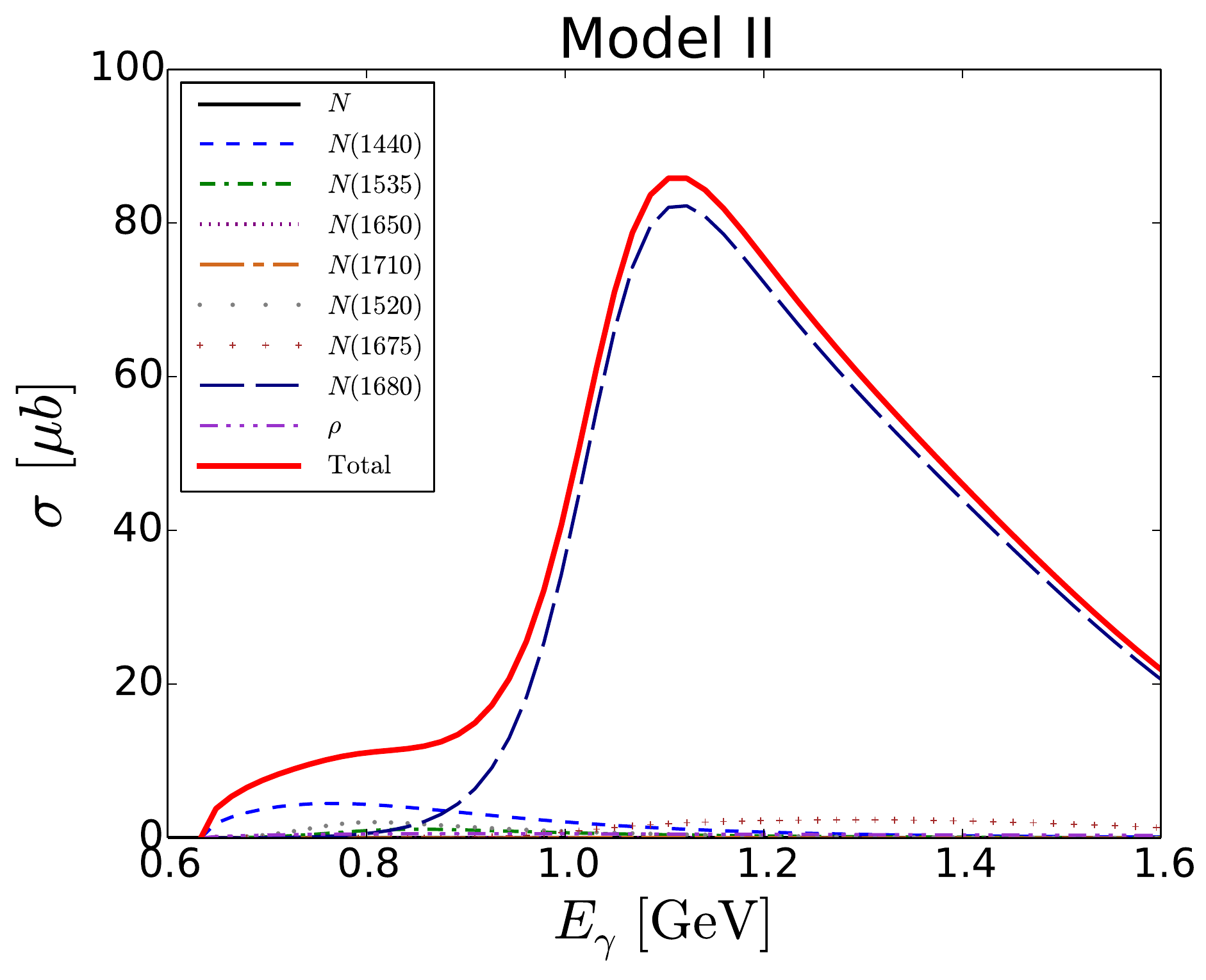}}
\caption{Total cross sections for the $\gamma N \to f_0(500)N$
  reaction as a function of $E_\gamma$. In the left panel, the results
from Model I are drawn, whereas in the right panel those from Model II
are depicted. Each contribution is distinguished by different types of
the curves.}
\label{fig:4}
\end{figure}
We need to delve into the physical reasons for Model I.
The $N(1440)1/2^+$ and $N(1535)1/2^-$ resonances increase the total
cross section of the $\gamma N\to f_0(500) N$ reaction near threshold.
As shown in the left panel of Fig.~\ref{fig:4}, the most dominant
contribution arises from the $N(1440)1/2^+$ resonance. Since the
$N(1650)1/2^-$ and $N(1710)1/2^+$ resonances have relatively smaller strong
coupling constants as well as photocouplings as listed in Tables~\ref{tab:3}
and \ref{tab:4}, the effect of these resonances is tiny. 
Interestingly, the contribution of the $\rho$ Reggeon in the $t$ channel
is rather suppressed. The effect of $N$ exchanges in the $s$ and $u$
channels is much smaller than that of $\rho$-Reggeon exchange through
the whole energy region. The magnitude of the total cross section for
$f_0(500)$ photoproduction is about 40 times larger than that for
$f_0(980)$ production in the case of Model I.

The right panel of Fig.~\ref{fig:4} depicts the results of the total
cross section for the $\gamma N\to f_0(500) N$ reaction from Model
II, where the $N^*$ resonances with higher spins are added in the $s$
channel in addition. The $N(1680)5/2^+$ resonance in the $s$ channel
yields a remarkably large contribution to the total cross section of
$f_0(500)$ photoproduction, so that its magnitude reaches even about
$80\,\mu \mathrm{b}$ around $E_\gamma\approx 1.1$ GeV.
There is at least one clear reason for this large contribution of the
$N(1680)5/2^+$. Firstly, the photocouplings of the $N(1680)5/2^+$ are
very large, as shown in Table~\ref{tab:4}, which come from the large
values of the experimental data on the photon decay
amplitudes $A_{1/2}$ and $A_{3/2}$~\cite{PDG2016}.
The value of $f_{1\gamma NN(1680)}$ is even
about 32 times larger than that of $f_{1\gamma NN(1440)}$. Moreover,
the size of the strong coupling constant for the $f_0(500) N N(1680)$
vertex is comparable to that of $g_{f_0(500) NN(1440)}$. Note that
even though the value of $g_{f_0(500) N N(1675)}$ is rather large, the
$N(1675)5/2^-$ resonance has almost no effect
on the total cross section because of its negative parity.
The results are not so sensitive to variations of the
cut-off masses. In this regard, the contribution of the $N(1680)5/2^+$
resonance discussed here seems to be robust.
Thus, it would be indeed of great interest if one could justify
experimentally whether the $N(1680)5/2^+$ resonance plays such a
dominant role in describing the $\gamma N\to f_0(500) N$ reaction.

\begin{figure}[htp]
\centering
\includegraphics[scale=0.42]{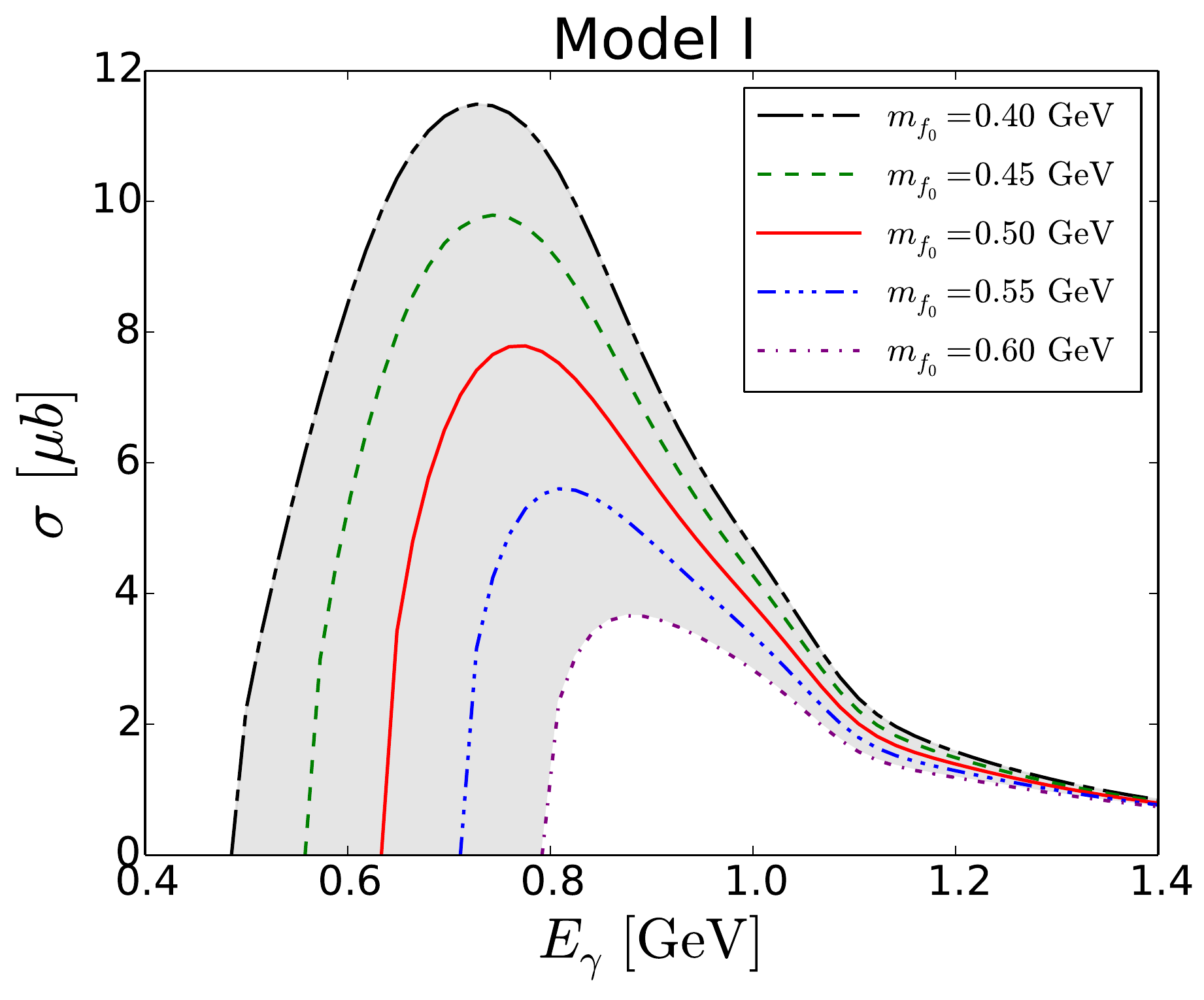}\hfill
\includegraphics[scale=0.42]{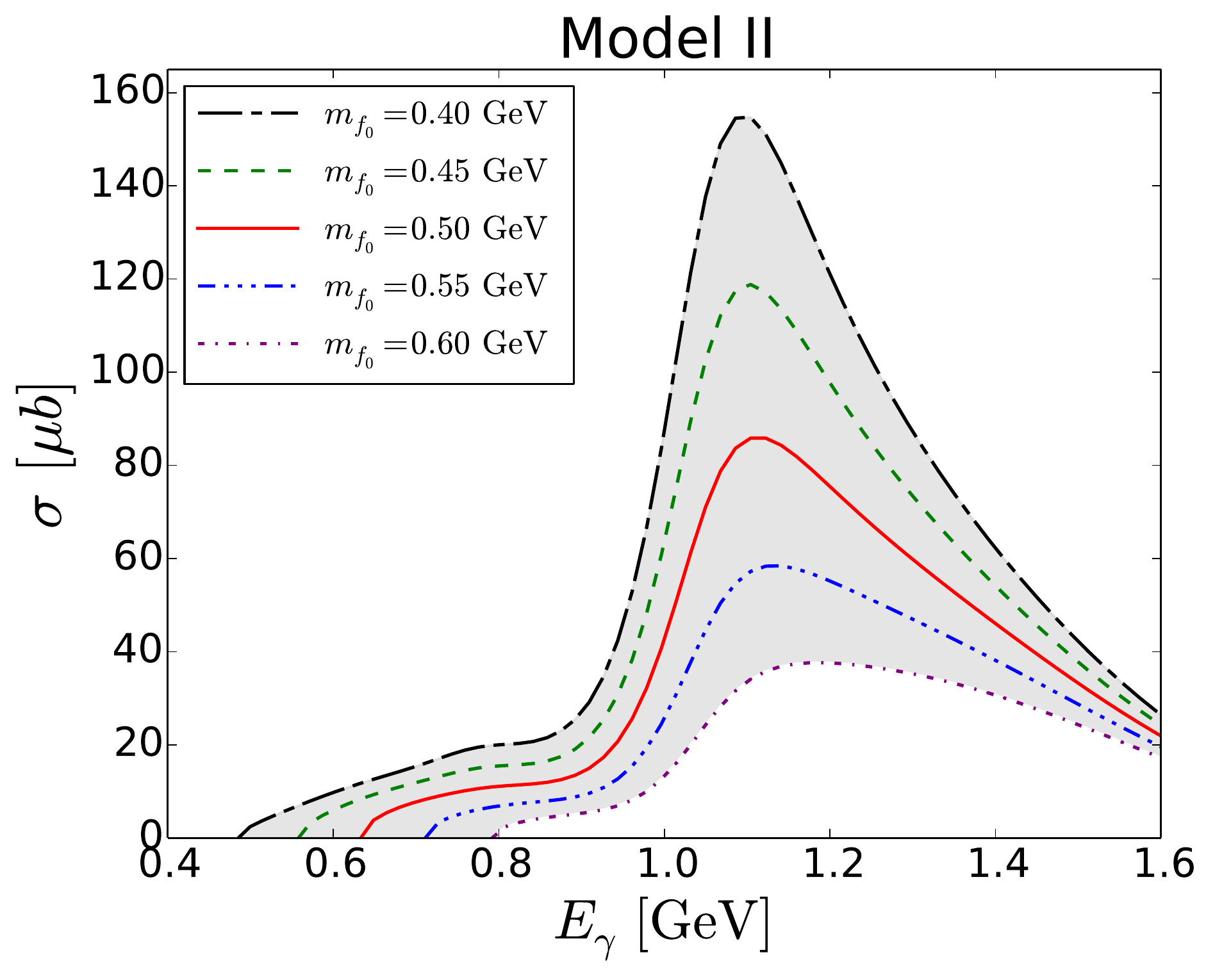}
\caption{$f_0(500)$ mass dependence of the total cross section for
the $\gamma N \to f_0(500)N$ reaction both in Model I and Model II.}
\label{fig:5}
\end{figure}
As mentioned briefly in Introduction, the uncertainty in the mass of
$f_0(500)$ is so large on account of its broad width, it is quite
unclear to settle the threshold energy. The PDG data has it that the
pole mass of $f_0(500)$ is $[(400-550)-i(200-350)]$
MeV~\cite{PDG2016}. In fact, the estimated mass of $f_0(500)$
is given in a wide range of its values, as listed in
Ref.~\cite{PDG2016}. Thus, we have to examine the dependence of
the total cross section on the mass of $f_0(500)$. In the left panel
of Fig.~\ref{fig:5}, we draw the results of the total cross section of
the $\gamma N\to f_0(500) N$ reaction from Model I with various values
of $m_{f_0}$ given between $0.4$ GeV and $0.6$ GeV. As expected, the
smaller values of $m_{f_0}$ produce the larger magnitudes of the total
cross section. Note that if one uses the value of $m_{f_0}$ larger
than 500 MeV, the $N(1440)1/2^+$ will be excluded because of the
larger threshold energy. Thus, the total cross section starts to get
reduced when the value of $m_{f_0}$ is larger than 500 MeV.
In the present work, we will take $m_{f_0(500)}=500$ MeV from now on.

\begin{figure}[htp]
\centering
\subfigure{\includegraphics[scale=0.42]{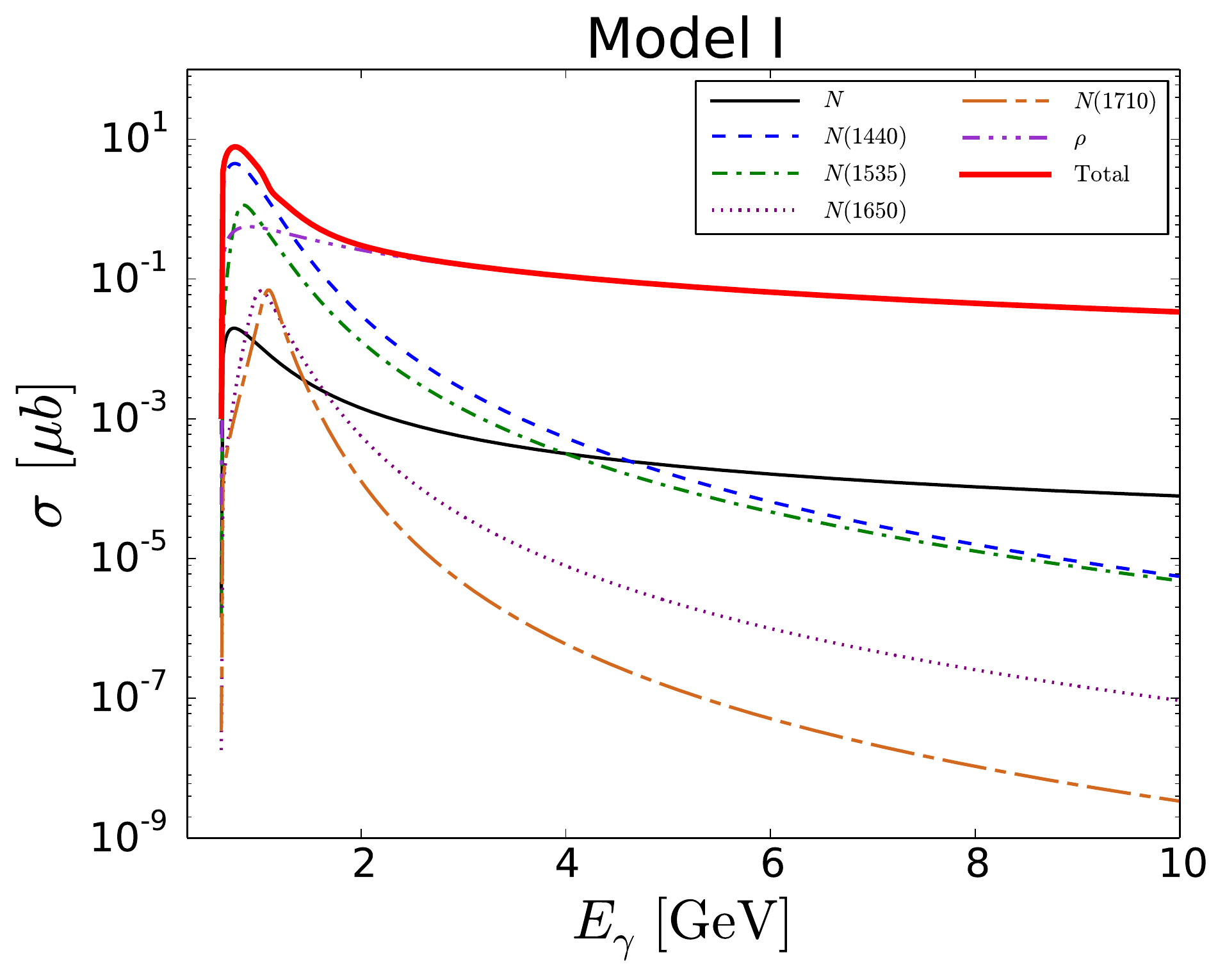}}\hfill
\subfigure{\includegraphics[scale=0.42]{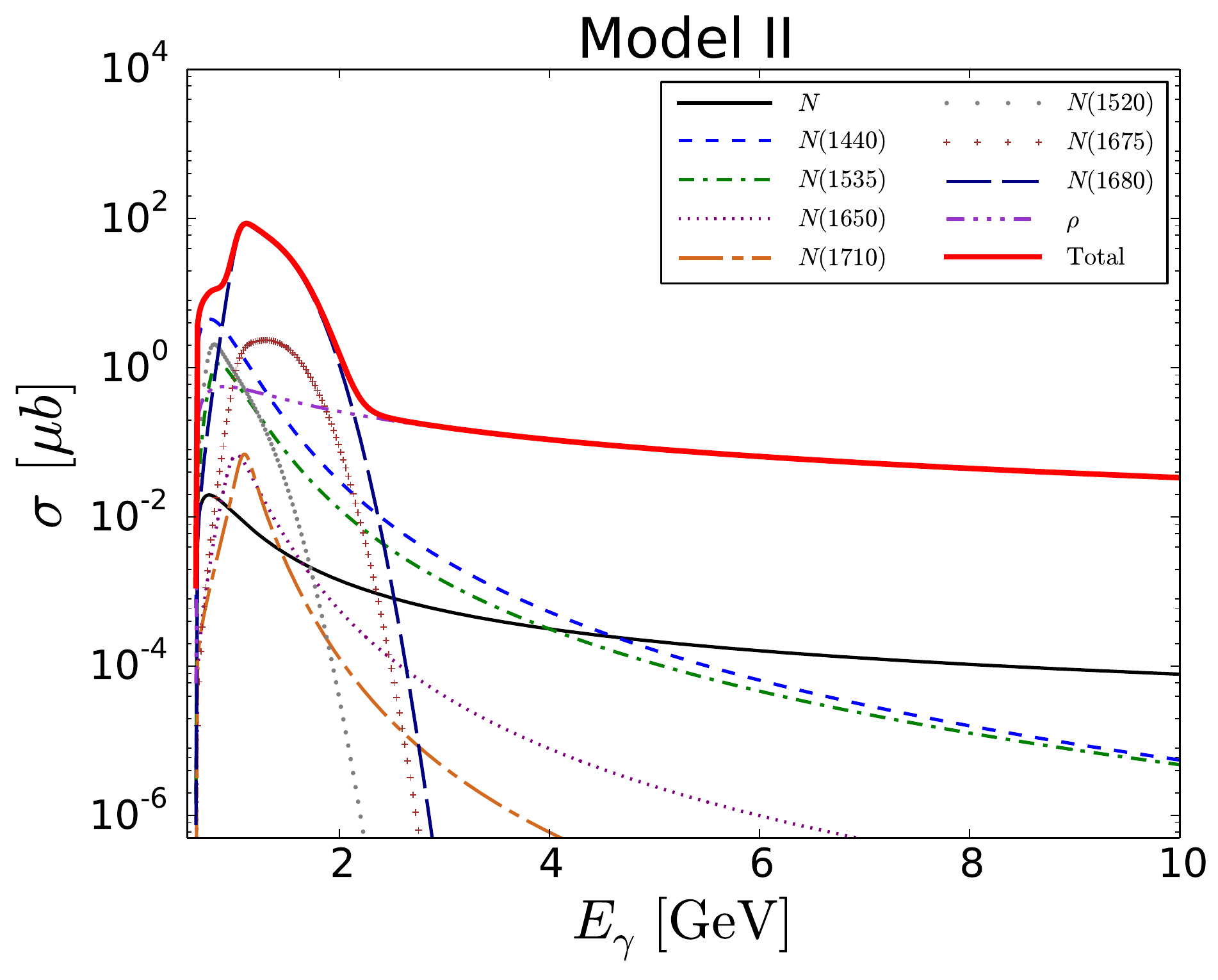}}
\caption{Total cross sections for the $\gamma N \to f_0(500)N$
  reaction as a function of $E_\gamma$ in the log scale . In the left
  panel, the results from Model I are drawn, whereas in the right
  panel those from Model II are depicted. Each contribution is
  distinguished by different types of the curves.}
\label{fig:6}
\end{figure}
In Figs.~\ref{fig:4} and \ref{fig:5}, we mainly have examined the
total cross section of $f_0(500)$ photoproduction in the vicinity of
the threshold energy. We now delve into the dependence of the total
cross section on $E_\gamma$ from the threshold energy through
$10$ GeV in the log scale. In the left panel of
Fig.~\ref{fig:6}, we show
the behavior of each contribution from Model I as $E_\gamma$
increases. As expected, all the resonance effects are diminished
quickly with the photon energy increased, while the
$\rho$-Reggeon in the $t$ channel takes over the contributions of all
the $N^*$ resonances around $E_\gamma\approx 2$
GeV and then dictates the dependence of the total cross section on
$E_\gamma$.
The $t$-channel Reggeon ensures the unitarity of the total cross
section, as shown in Fig.~\ref{fig:6}.
In the limit of $s\to \infty$, the unpolarized sum
of the Regge amplitude complies with the following asymptotic
behavior
\begin{align}
\lim_{s\to \infty} \sum_{\mathrm{pol}} |\mathcal{M}_t^{\mathrm{Regge}} (s,t)|^2
\varpropto s^{2\alpha_\rho(t)}.
\label{eq:asympt}
\end{align}
Thus, if one further increases $E_\gamma$, the contribution of the
$\rho$ meson startes to decrease, satisfying the asymptotic behavior
of Eq.~(\ref{eq:asympt}).
The right panel of Fig.~\ref{fig:6} shows the
results from Model II. Similarly, the contributions of most $N^*$
resonances fall off very fast as $E_\gamma$ increases. The effect of
the $N(1680)5/2^+$ lessens continuously after $E_\gamma\approx 1.2$
GeV, and then becomes smaller than those of the $\rho$
Reggeons around $E_\gamma \approx 2$ GeV. Thus, the $N^*$ resonances
come into play only near the threshold region as anticipated.
\begin{figure}[htp]
\centering
\subfigure{\includegraphics[scale=0.15]{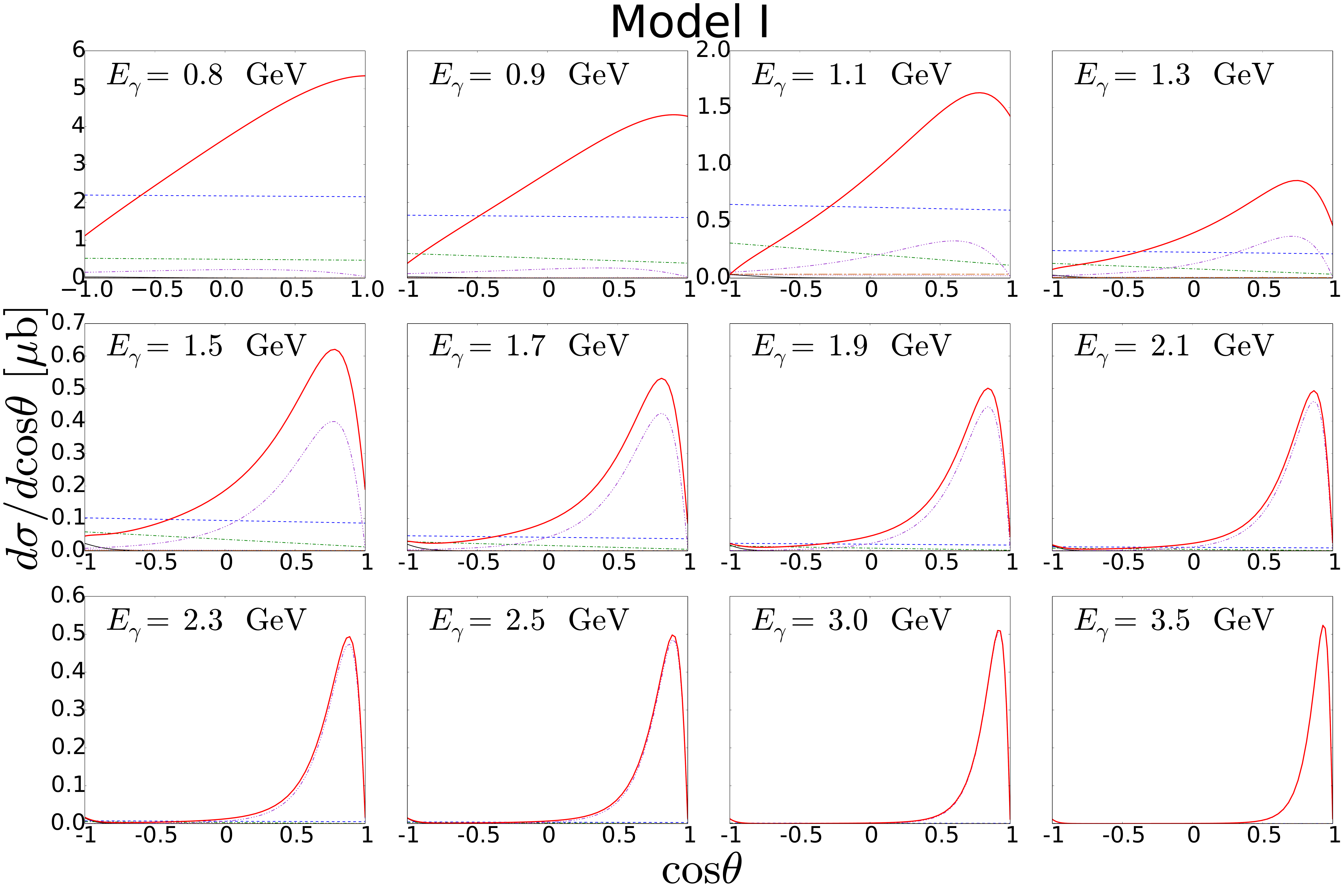}}\hfill
\subfigure{\includegraphics[scale=0.15]{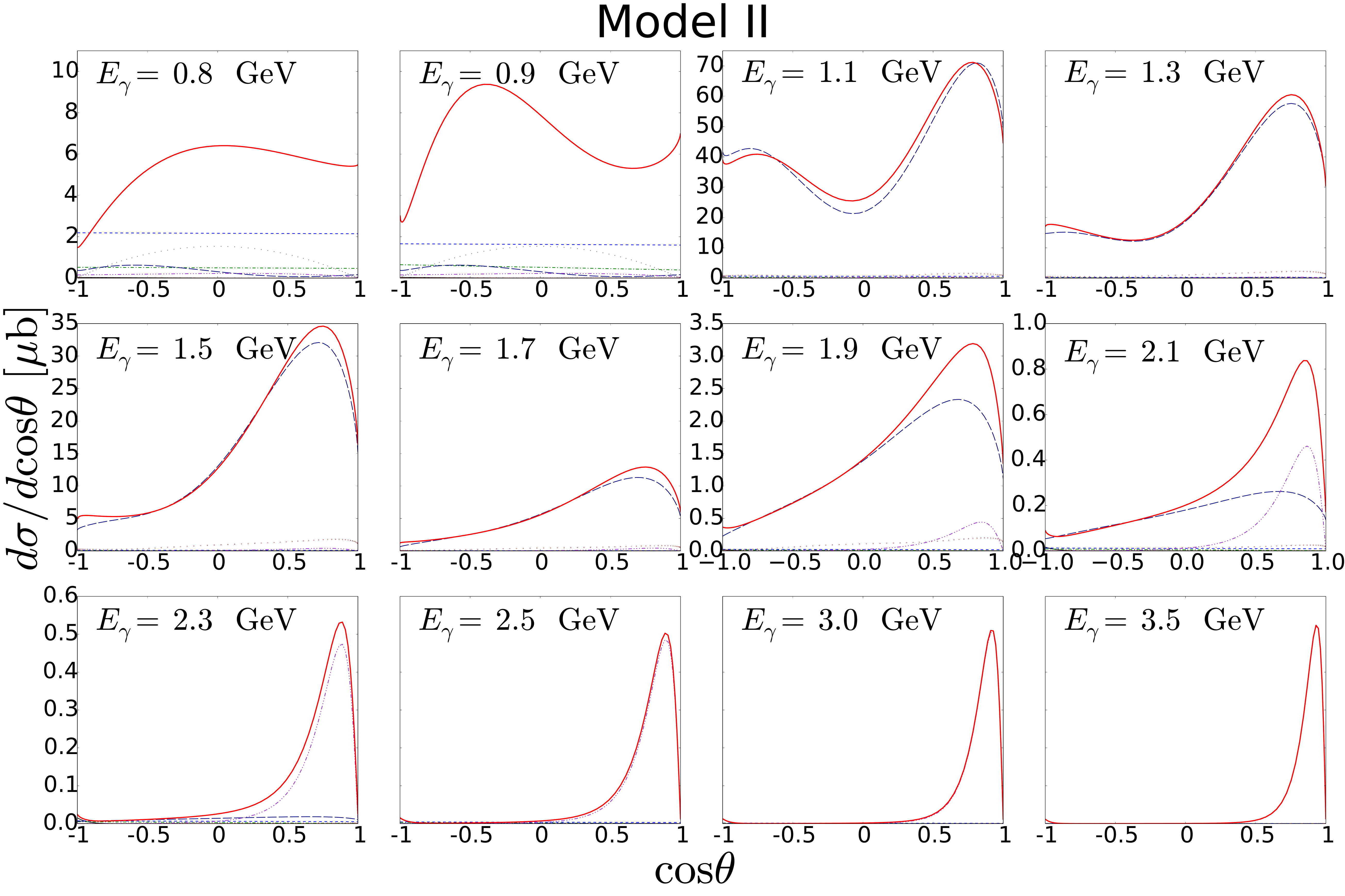}}
\caption{Differential cross section for the $\gamma N \to f_0(500)N$
  reaction as a function $\cos\theta$ with the photon energy
  $E_\gamma$ changed from 0.8 GeV to 3.5 GeV. The results from Model I
  are drawn in the upper panel, whereas those from Model II are
  depicted in the lower panel. Notations are the same as in
  Fig.~\ref{fig:4}.}
\label{fig:7}
\end{figure}

In the upper panel of Fig.~\ref{fig:7}, the results of the
differential cross section $d\sigma/d\cos\theta$ from Model I are
plotted as functions of $\cos\theta$, as the photon energy $E_\gamma$
is varied from 0.8 GeV through 3.5 GeV. Usually, the $s$-channel
contributions including all $N^*$ resonances with spin $1/2$ do not
show any $\cos\theta$ dependence. Note that Model I does not contain
any $N^*$ resonances with higher spins.
Nevertheless, the results of $d\sigma/d\cos\theta$ exhibit different
peculiarities. The differential cross section $d\sigma/d\cos\theta$ in
the forward region grows as $E_\gamma$ increases. However, the value
of $d\sigma/d\cos\theta$ almost vanishes at the very forward angle at
higher values of $E_\gamma$. This arises from
the structure of the amplitude of $\rho$ exchange in
Eq.~(\ref{Am_u}). 
Model II yields rather different results from those based on Model
I. Since the $N(1680)5/2^+$ resonance is the most dominant one from
the threshold energy through 1.5 GeV as shown in Fig.~\ref{fig:4},
and it has spin $5/2$ with positive parity, we expect that it will
have a certain effect on the $\cos\theta$ dependence of the
differential cross section. Indeed, it steers $d\sigma/d\cos\theta$ up
to $E_\gamma \approx 2.1$ GeV, as shown in the lower panel of
Fig.~\ref{fig:7}. As $E_\gamma$ increases more than
1.7 GeV, the $\rho$ Reggeon gains control of the $\cos\theta$
dependence of the differential cross section, so that we have more or
less the same results as in the upper panel of Fig.~\ref{fig:7}.

\begin{figure}[htp]
\centering
\subfigure{\includegraphics[scale=0.6]{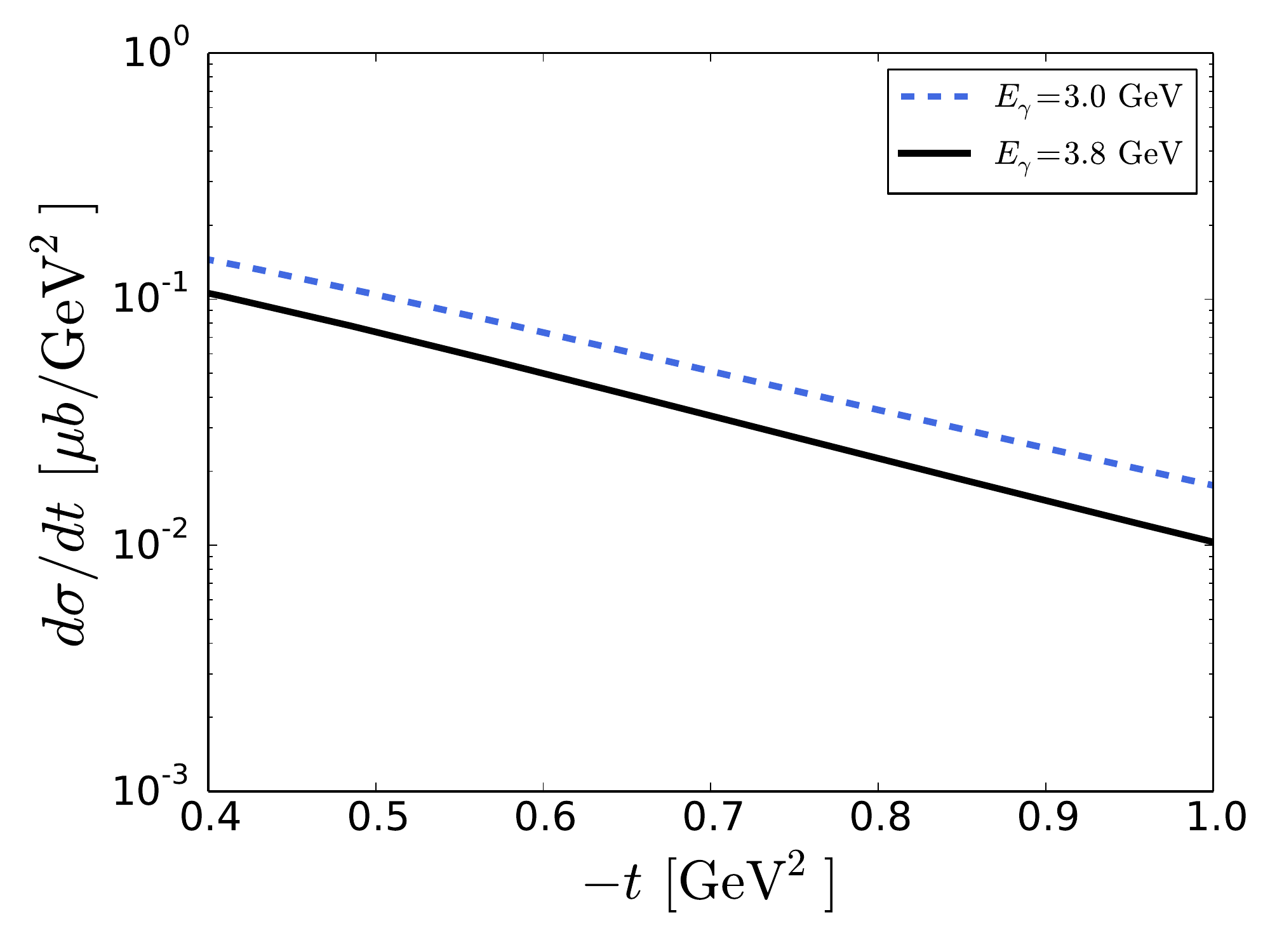}}
\caption{Differential cross section for the $\gamma N \to f_0(500)N$
  reaction as a function $t$ at two different photon energies,
  i.e. $E_\gamma=3.0$ GeV and $E_\gamma=3.8$ GeV. The solid curve
  represents the result at $E_\gamma=3.8$ GeV, whereas the dashed
  one plots that at $E_\gamma=3.0$ GeV.}
\label{fig:8}
\end{figure}
Figure~\ref{fig:8} draws the results of the differential cross section
$d\sigma/dt$ as a function of $-t$ in the range of $E_\gamma =
(3.0-3.8)$ GeV, so that we can directly compare them with those of
$f_0(980)$ photoproduction. The $t$ dependence is almost the same as
that of the $f_0(980)$ case, because we have exactly the same
$\rho$-Reggeon and $N$ exchanges. The only differences come from the
coupling constants. The contributions of the $N^*$ resonances are all
suppressed in this region of the photon energies. The magnitude of
$d\sigma/dt$ is approximately 2 times larger than that of
$f_0(980)$ photoproduction on account of different coupling constants.
Note that the differential cross section $d\sigma/dt$ obeys the
following asymptotic behavior
\begin{align}
\lim_{s\to \infty} \frac{d\sigma}{dt} (t\to 0) \varpropto
s^{2\alpha_\rho(0)-2}
\label{eq:asymptot2}
\end{align}
and the result shown in Fig.~\ref{fig:8} satisfies
Eq.~(\ref{eq:asymptot2}).

\section{Summary and conclusion}
We aimed in this work at investigating $f_0(500)$ and $f_0(980)$
photoproduction, based on a hybridized Regge model.
We first described the differential cross section $d\sigma/dt$ for the
$\gamma N\to f_0(980) N$ reaction, compared with the recent
experimental data on it. We fixed the relevant Regge parameters by
reproducing the data.
We introduced the $N^*$ contribution in the $s$
channel to study the production mechanism of the $\gamma N\to f_0(500)
N$ reaction. Since its threshold energy is much smaller than
$f_0(980)$ photoproduction, there exist several $N^*$ resonances that
can decay into $(\pi\pi)_{S-\mathrm{wave}}^{I=0}N$. Assuming that
$f_0(500)$ is much stronger than the background of the
$(\pi\pi)_{S-\mathrm{wave}}^{I=0}$ channel, we were able to find the
strong coupling constants for the $f_0(500) N N^*$ vertices. The
photocouplings of the $N^*$ resonances were determined by using the
experimental data on the corresponding photon decay amplitudes.
The cut-off masses for the form factors were fixed to be 1.0 GeV to
avoid additional ambiguity. In dealing with these $N^*$ resonances, we
constructed Model I and Model II. Model I included those with spin 1/2
only, while Model II was built in such a way that more $N^*$
resonances with higher spins were added to Model I. Near threshold, we
found that the $N(1440)1/2^+$ and $N(1535)1/2^-$ were dominant ones in
Model I, whereas the effects of other $N^*$ resonances
were almost negligible. In Model II, the contribution of the
$N(1680)5/2^+$ dictates the total cross section of the
$\gamma N\to f_0(500) N$ reaction. Remarkably, the $N(1680)5/2^+$
resonance enhances the magnitude of the total cross section up to about
$80\,\mu\mathrm{b}$. The main reason comes from the large value of its
photoncouplings.  The strong coupling constant of the $N(1680)5/2^+$
is also relatively large.
Since the mass of the $f_0(500)$ meson is not precisely fixed because
of its large width, we examined the dependence of the total cross
section on its mass in the range of $(400-600)$ MeV. As expected,
small the $f_0(500)$ mass was, the larger the total cross section
was in Model I and Model II.
If $E_\gamma$ increases, then the $N^*$ contribution fade away very
fast, so that $\rho$-Reggeon exchange takes over the control as in
the case of $f_0(980)$ photoproduction.
We also computed the differential cross section $d\sigma/d\cos
\theta$. While the contributions of the $N^*$ resonances in the $s$
channel are rather flat in the case of Model I,  $N(1680)5/2^+$
governs $\cos\theta$ dependence again because of its high spin.
Finally, we computed the differential cross section $d\sigma/dt$ for
the $\gamma N \to f_0(500) N$ reactoin. The results showed that the
$t$ dependence and the magnitude looked very similar to those for
$f_0(980)$ photoproduction.

Though it is very difficult to study $f_0(500)$ photoproduction
experimentally, it is still of great importance to study the
production mechanism of the $\gamma N\to f_0(500) N$ reaction, since
it cast light on the structure of the $N^*$ resonances as investigated
in the present work. It also provides a certain clue in studying more
complicated processes with three-particle final states such as $\gamma
N\to (\pi\pi)_{S-\mathrm{wave}}^{I=0} N$ in the future.

\begin{acknowledgments}
H-Ch.K. is grateful to M. V. Polyakov and the members of TPII in
Ruhr-Universit\"at Bochum for valuable discussions and hospitality
during his visit, where part of the work has been carried out. The
present work was supported by Basic Science Research Program through
the National Research Foundation of Korea funded by the Ministry of
Education, Science and Technology (Grant Number:
NRF-2015R1D1A1A01060707). S.H.K acknowledges support from the Young
Scientist Training Program at the Asia Pacific Center for Theoretical
Physics by the Korea Ministry of Education, Science, and Technology,
Gyeongsangbuk-Do and Pohang City.
\end{acknowledgments}

\end{document}